\documentclass[11pt]{article}

\usepackage{amsthm,color,url,graphicx,pstricks,pst-node,pst-plot,pst-coil,amsmath,amscd,bm,pifont,xspace,amssymb,wrapfig,array,ctable}
\usepackage[numbers,sort&compress]{natbib}

\newtheorem{Theorem}{Theorem}
\newtheorem{Lemma}[Theorem]{Lemma}
\newtheorem{Corollary}[Theorem]{Corollary}

\newtheorem{Proposition}{Proposition}
\newtheorem{fact}{Fact}

\newcommand{\IE}{{\em i.e.}\xspace}
\newcommand{\EG}{{\em e.g.}\xspace}

\newcommand{\cB}{\mathcal{B}}

\newcommand{\cP}{\mathcal{P}}

\newcommand{\cE}{\mathcal{E}}
\newcommand{\eps}{\varepsilon}
\newcommand{\cL}{\mathcal{L}}
\newcommand{\R}{{\mathbb R}}  
\newcommand{\Ex}{{\mathbb E}} 
\newcommand{\K}{{\mathcal K}}

\newcommand{\F}{\mathsf{F}}

\newcommand{\G}{{\mathcal G}}

\newcommand{\x}{\mathbf{x}}
\newcommand{\NP}{\mathsf{NP}}
\newcommand{\poly}{\mathsf{poly}}
\newcommand{\QIP}{{\sf QIP}}

\newcommand{\balpha}{{\bm \alpha}}

\renewcommand{\v}{\mathcal{V}}
\renewcommand{\u}{\mathcal{U}}

\newcommand{\comment}[1]{}
\newcommand{\bee}{\mathbf{e}}
\newcommand{\SBO}{{\sc Sbo}}

\newcommand{\SSBO}{{\sc Ssbo}}
\newcommand{\DSSBO}{{\sc Dual-Ssbo}}
\newcommand{\USSBO}{{\sc Uniform-Ssbo}}
\newcommand{\UFSSBO}{{\sc Uniform-Frac-Ssbo}}
\newcommand{\UISSBO}{{\sc Uniform-Int-Ssbo}}
\newcommand{\ISSBO}{{\sc Int-Ssbo}}
\newcommand{\DISSBO}{{\sc Dual-Int-Ssbo}}
\newcommand{\FSSBO}{{\sc Frac-Ssbo}}
\newcommand{\DFSSBO}{{\sc Dual-Frac-Ssbo}}
\newcommand{\MSSBO}{{\sc Multi-Ssbo}}
\newcommand{\DMSSBO}{{\sc Dual-Multi-Ssbo}}
\newcommand{\IMSSBO}{{\sc Int-Multi-Ssbo}}

\newcommand{\FMSSBO}{{\sc Frac-Multi-Ssbo}}

\newcommand{\Epayoff}{{\Ex[\mbox{payoff}]}}

\begin{document}

\title{Stochastic Budget Optimization in Internet Advertising}

\author{
Bhaskar DasGupta \\
Department of Computer Science \\
University of Illinois at Chicago \\
Chicago, IL 60607 \\
{\tt dasgupta@cs.uic.edu}
\and
S. Muthukrishnan \\
Department of Computer Science \\
Rutgers University \\
Piscataway,  NJ 08854 \\
{\tt muthu@cs.rutgers.edu}
}

\maketitle

\begin{abstract}
Internet advertising is a sophisticated game in which the many advertisers ``play'' to optimize their return on investment. 
There are many ``targets'' for the advertisements, and each ``target'' has a collection of games with a potentially
different set of players involved. In this paper, we study the problem of how advertisers allocate their budget across these ``targets''. 
In particular, we focus on formulating their {\em best response} strategy as an optimization problem. 
Advertisers have a set of keywords (``targets'') and some stochastic information
about the future, namely a probability distribution over {\em scenarios} of cost vs click
combinations. This summarizes the potential states of the world assuming that the strategies of other players are fixed. 
Then, the best response can be abstracted as {\em stochastic budget optimization}
problems to figure out how to spread a given budget across
these keywords to maximize the expected number of clicks. 

We present the {\em first known} non-trivial poly-logarithmic approximation for
these problems as well as the first known hardness results of getting better than logarithmic approximation ratios in the various parameters involved.
We also identify several special cases of these problems of practical interest, such as 
with fixed number of scenarios or with polynomial-sized parameters related to cost,  
which are solvable either in polynomial time or with improved approximation ratios. 
Stochastic budget optimization with scenarios has sophisticated technical structure. Our approximation and hardness results come from relating these 
problems to a special type of ($0/1$, bipartite) quadratic programs inherent in them.
Our research answers some open problems raised by 
the authors in ({\sf Stochastic Models for Budget Optimization in Search-Based Advertising, Algorithmica, 58 (4), 1022-1044, 2010}). 
\end{abstract}

\section{Introduction}

This paper deals with the problem of how advertisers allocate their budget in Internet advertising. 
In sponsored search, users who pose queries to internet search engines are not only provided search
results, but also a small set of text ads. These ads are chosen from a set of campaigns set up by
advertisers based on the keywords in the search query. A lot of focus has been on how these ads are chosen and 
priced, which is via an auction that is by now well known~\cite{Varian,EOS,AGM}\footnote{Likewise, there was a lot of work on 
bidding strategies~\cite{borgs,Chak,FMPS07,MPS07}. This paper extends that body of work by considering a 
richer model of uncertainty; see subsequent paragraphs.}.
Our focus is instead 
on the problem faced by advertisers. Even small advertisers have many {\em keywords}, a budget in mind and must figure out how to spread this budget on bids 
for each of these keywords. This is a highly nontrivial
task, and the basis for a separate industry to support advertisers. 
A similar problem arises with ``display ads'' where advertisers have {\em websites} where their
ads will be shown and need to split their 
budget for the ad campaign across the sites to be most effective. Likewise, in 
behavioral targeting, advertisers have to decide how to spread their budget across {\em behavior groups}. 
In all these cases, therefore, advertisers have various ``targets'' and wish to split their budget across them to optimize their ad campaigns. 

Consider the sponsored search example and fix an advertiser $A$. They have many keywords that they would like to target for their
ads. How should they bid for each, given some overall budget they can spend? 
There is a sophisticated underlying game in which the many advertisers ``play'' to optimize their return on investment
simultaneously. For each keyword and for each instance of auction triggered by this keyword, there is potentially 
a different set of competing advertisers involved. Building effective strategies is challenging amidst so many parameters. A fundamental and widely 
accepted proposal is for the advertiser $A$ to pursue a {\em best response} strategy, \IE, fix the strategies of other advertisers and pick the 
best strategy as one's response. Besides being a simple and easy strategy to understand and hence suitable for experimentation by advertisers, best response has desirable
properties. For example, in the absence of budgets and for single repeated auctions, special type of best response by every player leads to the 
VCG outcome~\cite{c1,c2,Varian,EOS}. 

In order to help the advertisers implement this best response strategy, 
search engines provide them with expected bid versus clicks function for each 
keyword\footnote{See, for example, 
Traffic Estimator at~\url{http://adwords.google.com/support/aw/bin/answer.py?hl=en&answer=8692}, 
bidding tutorial at~\url{http://adwords.google.com/support/aw/bin/answer.py?hl=en&answer=163828} and 
bid simulator at~\url{http://adwords.google.com/support/aw/bin/answer.py?hl=en&answer=138148}}. 
Assuming that the rest of the world is fixed, these functions provide an estimate of the expected number of 
clicks an advertiser would obtain by bidding a certain value on that keyword. These functions
can also be ``learned'' by an advertiser to some extent by systematically trying out various bids. 
Finding advertiser's best response bidding strategy then becomes an optimization problem where the goal is to maximize the 
expected number of clicks assuming access to these functions. 
The resulting problems are in the spirit of the Knapsack problem~\cite{borgs,Chak,FMPS07,MPS07} with 
many of them solvable nearly exactly or with constant factor approximations.

A more general approach is to acknowledge that, in reality, the bids vs clicks functions are {\em not fixed}, but 
rather random variables with unknown correlations and uncertainties: number of queries (and hence, clicks and 
budget spent on a keyword) change each day, relative occurrences of keywords change (\EG, searches for
beach and snow are complementary\footnote{See \url{www.google.com/trends?q=beach\%2C+snow&ctab=0&geo=all&date=all&sort=0} for yearly
and \url{www.google.com/trends?q=clubs\%2C+stocks&ctab=0&geo=all&date=mtd&sort=0} for weekly 
trends.}), and so on.
Therefore, one has to consider a 
specific stochastic model for these random variables and then maximize the {\em expected number of clicks} under that model. 
This approach was initiated in~\cite{MPS07} leading to 
a stochastic budget optimization problem that is studied in this paper.

\subsection{Organization of the paper}

For convenience of the readers, we organize the rest of the paper in the following manner.
\begin{itemize}
\item
We start with Section~\ref{model} which describes all of our stochastic budget optimization models and corresponding computational 
problems precisely, starting from the simplest one, together with some comments and justifications about the model.
In the last subsection of this section (Section~\ref{sec-notations}), 
we fix some notational uniformity for readers convenience.

\item
In Section~\ref{sec-results}, we summarize the results obtained in this paper. For the benefit of the reader, we 
group the results into two categories, namely a set of main results that deal with the computational complexity issues of the 
original models without restrictions and a set of additional results that deal with variations and special cases of the models
defined in Section~\ref{model}.
\end{itemize}
The remaining sections of the paper, excluding conclusion and references, deal with precise statements of our results 
and technical details of their proofs.
For complex proofs, we first provide a more informal overview of the steps in the proof before proceeding with
technical details. These sections are organized in the following manner.
\begin{itemize}
\item
In Section~\ref{whyqp} we discuss the quadratic integer programming reformulations of the various \SBO\ problems.

\item
In Section~\ref{sec-polylog} we state and prove our 
poly-logarithmic approximation algorithms for \SSBO\ and \MSSBO\ problems (main result {\bf (R1)}).

\item
In Section~\ref{hardnes}, we state and prove our 
approximation-hardness results for both \SSBO\ and \MSSBO\ problems (main result {\bf (R2)}).

\item
Section~\ref{other-results} contain all other results: 
\begin{itemize}
\item
In Section~\ref{special-cases} 
we show that many \SSBO\ problems have improved solutions if certain parameters are restricted 
in their range of values.

\item
In Section~\ref{a1} we show the limitations of semidefinite programming based approaches for solving \SSBO\ 
problems.
\end{itemize}
\end{itemize}

\section{Scenario Model for Stochastic Budget Optimization}
\label{model}

We discuss the scenario model\footnote{{The scenario model was introduced in~\cite{MPS07}.
For a very detailed discussion of prior works related to the approach in the model, see Section~$1.4$ of~\cite{MPS07}}.}
and related problems using the language of sponsored search\footnote{Our discussion can easily be 
adopted to other internet ad channels like display ads and behavioral targeting.}.
We use the suffix \SSBO\ (Scenario Stochastic Budget Optimization) for various acronyms for 
different versions of our problems.
For the convenience of the readers and to delay introducing more involved notations, we first
start with a slightly simpler version of the model involving only one slot. We refer to this version as 
the ``uniform cost'' case and describe it in the next section.

\subsection{Single Slot Case: Uniform Cost Model}
\label{sec-singleslot}

This basic model starts with the following assumptions:
\begin{itemize}
\item
There is a single slot for advertising.

\item
We have a set of $n$ keywords $\K_1,\K_2,\ldots,\K_n$ with 
the keyword $\K_j$ having a 
{\em cost-per-click} $d_j$ (a positive integer).

\item
We have a positive integer $B$ denoting the {\em budget} for the advertiser.

\item
We have a collection of $m$ ``scenarios'' where the $i$th scenario is 
characterized by the following parameters: 
\begin{itemize}
\item
A probability of $\eps_i$ ($\sum_{i=1}^m\eps_i=1$).
\item
A ``click vector'' $(a_{i,1},a_{i,2},\ldots,a_{i,n})$
where each $a_{i,j}\geq 0$ is an integer.
Each $a_{i,j}$ denotes the number of clicks obtained by the $j$th
keyword $\K_j$ in the $i$th scenario.
\end{itemize}
\end{itemize}
Scenarios can be thought of as sampling the model over various times\footnote{Scenarios can be provided by 
the search engine for the advertisers, or used by the search engines to bid on behalf of 
advertisers. Similarly, advertisers and other search engine optimizers can also ``infer'' 
scenarios indirectly using trends and other data provided by search engines.}. 

Our general goal is to compute $n$ selection variables $x_1,x_2,\ldots,x_n$,
where $x_j$ corresponds to the $j$th keyword, to {\em maximize} a suitable total payoff.
A crucial aspect of the discussed 
formulation is that, if the budget is not limiting, then the payoff corresponds to
the total number of expected clicks, but if the budget turns out to be limiting for any scenario then
the payoff {\em scales} the total number of expected clicks by the fraction that the budget
would provide\footnote{The underlying assumption is that, within a scenario, the queries and keywords are 
well-mixed and, when budget runs out, the ad campaign is halted for the period as is currently
done. The queries and keywords are well-mixed not only because of aggregation of streams from millions of
users but also because of ad throttling that spreads out the eligible ad campaigns over the period of a scenario. 
See~\cite{MPS07} for exact details of justification.}.
Based on the above intuition, our precise goal is {\em maximize} the total {\em expected} payoff 
over {\em all} scenarios, \IE, 
\[
\mbox{\sf maximize } \Ex[\mbox{payoff}]=\sum_{i=1}^m \Ex[\mbox{payoff}_i] 
\]
where
the \textsf{expected payoff} 
$\Ex[\mbox{payoff}_i]$ 
for the $i$th scenario is 
\begin{gather}\label{eq1}
\Ex[\mbox{payoff}_i]=\left\{
\begin{array}{cl}
\eps_i \sum_{j=1}^n a_{i,j} x_j, & \mbox{   if $\sum_{j=1}^n a_{i,j} d_{j} x_j \leq B$} \\ 
\dfrac{B}{\sum_{j=1}^n a_{i,j} d_{j} x_j}\, \left(\eps_i\sum_{j=1}^n a_{i,j} x_j \right), & \mbox{   otherwise} \\
\end{array}
\right.
\end{gather}
Following~\cite{MPS07}, we distinguish between two versions of the problem based 
on the nature of the selection variables:
\begin{description}
\item[Integral version (\UISSBO):] 
$x_j\in\{0,1\}$ for all $j$.
This corresponds to the case when based on the stochastic information, either the advertiser
chooses to win and pay for all clicks for a keyword, or not at all. Hence, the strategy of the advertiser is {\em deterministic}.

\item[Fractional version (\UFSSBO):] 
$0\leq x_j\leq 1$ for all $j$.
This can be thought of as a strategy in which the advertiser treats these numbers as
{\em probabilities} and bids for the keywords in a {\em randomized} fashion based on these probabilities,
thereby only winning (and paying for) a portion of all clicks and impressions for each keyword. 
If the deterministic strategy is hard to compute and provides a solution of bad quality
then the randomized strategy is more desirable.
\end{description}
Other than the scenario model, there are at least two other possible models
for stochastic budget optimization as discussed in~\cite{MPS07}.
In the {\em proportional model} there is just one global random variable for the total number of 
clicks in the day that keeps the relative proportions of clicks for different keywords the same, whereas
in the {\em independent keywords model} each keyword comes with its own probability distribution.
However, among all these models this scenario-based model is perhaps one of the most natural model
of reality and provides an appropriate middle ground between complex arbitrary joint probability distribution and
a single distribution for all keywords. 
It was shown in~\cite{MPS07} that both \UISSBO\ and \UFSSBO\ are $\NP$-hard.
In the sequel, we assume without loss of generality 
that $1=d_1\leq d_2\leq\ldots\leq d_n$.

\subsection{Single Slot Case: General Model}
\label{stretch}

In a more realistic version of the \SSBO\ problems the cost-per-click values may
vary {\em slightly} over a range of scenarios due to their small errors in estimation. 
This can be modeled by introducing a stretch parameter ({\em small integer})\footnote{Throughout the 
paper, the notation $\poly(a)$ denotes a polynomial in $a$, \IE, $a^c$ for some positive constant $c$.} 
$1\leq\kappa=O\left(\poly(\log (m+n))\right)$.
Now, $d_j$ stands for the {\em basic} cost-per-click for the keyword $\K_j$, whereas the 
{\em real} cost-per-click for the keyword $\K_j$ in the $i$th scenario is denoted by 
$c_{i,j}$, with $c_{i,j}\in [d_j,\kappa d_j)$\footnote{For example, 
the stretch parameter $\kappa$ allows us to model situations such as when
the real costs can be drawn from a probability distribution 
with a mean around $\frac{1+\kappa}{2}d_j$ with a negligible probability
of occurring outside a range of $\pm\frac{1-\kappa}{2}d_j$ of the mean. Note that this is just an illustration.
We do not assume any specific probability distribution for the variations of the real costs per click 
except that it varies within an interval of length $\kappa$.}. 
Then, Equation~\eqref{eq1} can be simply updated by replacing 
$d_j$ in the equation of the $i$th scenario by $c_{i,j}$.
We refer to the integral and fractional versions of this general case
as \ISSBO\ and \FSSBO, respectively; note that the \USSBO\ problems
are obtained from the corresponding \SSBO\ problems by setting $\kappa=1$.

\subsection{Multi Slot Model}
\label{sec-multislot}

In the multi-slot case there are $s\geq 1$ {\em slots} for each keyword with 
the {\em generalized second price auction} for these slots.
Let $d_{j,k}$ be an integer denoting the value of the {\em basic} cost-per-click 
the $k$th slot of the $j$th keyword; we assume 
$d_{j,1}\leq d_{j,2}\leq\cdots\leq d_{j,s}$.
Let $c_{i,j,k}\in [d_{j,k},\kappa d_{j,k})$
denote the value of the {\em real} cost-per-click 
for the $k$th slot of the $j$th keyword in the $i$th scenario
where $\kappa$ is the stretch parameter as in Section~\ref{stretch}, 
and let $B>0$ denote the {\em budget} (a positive integer) for the advertiser.
Our goal is now to compute a set of $sn$ selection variables $x_{j,k}$ 
where the selection variable $x_{j,k}$ corresponds to $k$th slot for the $j$th keyword.
We again have a collection of $m$ scenarios where the $i$th scenario is
characterized via: 
\begin{itemize}
\item
a probability $\eps_i$ ($\sum_{i=1}^m\eps_i=1$), and 

\item
a ``click vector'' $(a_{i,j,1},a_{i,j,2},\ldots,a_{i,j,s})$
where each $a_{i,j,k}$ is a non-negative integer denoting 
the number of clicks obtained by the $k$th slot of the $j$th keyword $\K_j$ in the $i$th scenario.
\end{itemize}
The goal is to compute the allocation variables $x_{j,k}$'s with the constraints
\begin{equation}\label{multislot-constraint} 
\forall\, j\,\colon\,\sum_{k=1}^sx_{j,k}\leq 1
\end{equation}
to maximize 
the total expected payoff 
\[
\Ex[\mbox{payoff}]=\sum_{i=1}^m\Ex[\mbox{payoff}_i]
\]
where
\begin{gather}
\Ex[\mbox{payoff}_i]=\left\{
\begin{array}{cl}
\eps_i \sum_{j}\sum_{k} a_{i,j,k} x_{j,k}, & \mbox{if $\sum_{j}\sum_{k} a_{i,j,k} c_{i,j,k} x_{j,k} \leq B$} \\
\dfrac{B}{\sum_{j}\sum_{k} a_{i,j,k} c_{i,j,k} x_{j,k}}\left( \eps_i\sum_{j}\sum_{k} a_{i,j,k} x_{j,k}\right), & \mbox{otherwise} \\
\end{array}
\right.
\end{gather}
We again distinguish between two versions of the problem:
\begin{description}
\item[Integral version (\IMSSBO):]
$x_{j,k}\in\{0,1\}$ for all $j$ and $k$.
Here, $x_{j,k}=1$ if the advertiser selects the $k$th slot
for the $j$th keyword, and  $x_{j,k}=0$ otherwise.

\item[Fractional version (\FMSSBO):]
$0\leq x_{j,k}\leq 1$ for all $j$ and $k$.
Here, $x_{j,k}$ 
denotes the probability that the advertiser selects the $k$th slot
for the $j$th keyword and $1-\left(\sum_{k=1}^s x_{j,k}\right)$ is the probability with which
the advertiser does not bid on the $j$th keyword at all. 
\end{description}
Note that the scenario model for multi-slot stochastic budget optimization is {\em quite 
different} in nature from the other multi-slot models such as the one discussed in~\cite{FMPS07}
since, for example, one can go under or over the budget in one scenario to get a
better overall expected payoff.

\subsection{Relevance and Significance of Scenario Models}

Scenario models are a popular way of modeling optimization problems 
involving uncertainties in parameters by 
creating a number of scenarios that depict the probability distribution 
of various possibilities and then provide a solution that optimizes 
the expectations of outcomes over these scenarios. 
The scenario model is important for at least two reasons as explained in~\cite{MPS07}, which we state below. Firstly, market analysts often think of uncertainty by 
{\em explicitly} creating a set of a few model scenarios, possibly attaching a weight to each
scenario. Secondly, the scenario model gives us an important tool into understanding the fully general problem with arbitrary joint distributions. 
Allowing the full generality of an arbitrary joint distribution gives us significant modeling power, but poses challenges to the
algorithm designer. Since a naive explicit representation of the joint distribution requires space
exponential in the number of random variables, one often represents the distribution implicitly by
a sampling oracle. A common technique, Sampled Average Approximation, is to replace the
true distribution by a uniform or non-uniform distribution  over a set of samples drawn by some
process from the sampling oracle, effectively reducing the problem to the scenario model. 
In addition to their usual applications in operations research (\EG, see~\cite{D91}),
this approach is getting more and more attention in Wall Street as financial portfolios are being created in this way (\EG, see~\cite{Z96}).
For example, Cocco, Consiglio and Zenios in~\cite{CCZ00}  
developed a scenario-based optimization model for asset and liability management of participating insurance policies with minimum guarantees
and Mausser and Rosen in~\cite{HR01} 
developed three scenario optimization models for portfolio credit risk.

In sponsored search, this is an appropriate model
and embodies the ``best response'' strategy.
There is a complex function that maps the state of the world and the users to the queries they pose and their
actions such as whether they click on ads. The search engines give a limited amount of information to help advertisers\footnote{For
example, see \url{https://adwords.google.com/select/TrafficEstimatorSandbox}}, and advertisers can 
learn various scenarios that determine their click vs cost behaviors to some extent 
by running experiments, analyzing their web traffics {\em etc}\xspace. However, sponsored search products only provide a limited bidding language to 
structure one's campaign\footnote{See for example, \url{http://algo.research.googlepages.com/ec09-partI.pdf}} and hence, necessarily,
most advertisers have to target different scenarios simultaneously with each bidding choice. This is the stochastic budget optimization
problem we study in this paper. 
One natural idea is for advertisers to recognize in real time the particular 
scenario one faces and then apply the best bidding for that scenario. 
However, this is difficult to do in practice because of limited and delayed 
information in the system, and it is also expensive to implement. 
Furthermore, scenario models provide us with an important first step into understanding the fully general problem with 
arbitrary joint distributions that might be hard to model and analyze since, for example, naive explicit representation of a joint distribution may require space that is 
exponential in the number of random variables. Instead, techniques such as 
Sampled Average Approximation explained in the preceding paragraph are used,
effectively reducing the problem to the scenario model.
Thus, stochastic budget optimization problems under 
the scenario model are very appropriate for sponsored search applications. 

We do acknowledge that other strategies besides the ``best
response'' may be used by advertisers in practice\footnote{By other strategies, we mean strategies in which 
the advertiser does not fix the strategies of other advertisers.}, and stochastic budget optimization
algorithms proposed here are not currently used within the practical tools that
are publicly available. Nevertheless, best response is a
reasonable strategy (even recommended by some search engines), and indeed 
many anecdotal conversations with advertisers and sponsored search optimizers have
clearly indicated to us that they would like to bid to balance across
myriad of scenarios. 
Our algorithms in this paper (even the dynamic programming based ones) can be {\em easily} implemented in current systems. 

\subsection{Notational Remarks}
\label{sec-notations}

As the reader may have already observed, precise definitions of the various models involve a lot of
variables and subscripts. To make the exposition clearer, we will therefore adopt the following conventions:
\begin{itemize}
\item 
For variables involving keywords, scenarios and (for the multi-slot model) slots, we will 
use subscripts $i$, $j$ and $k$ (and their obvious variations such as $i_1$, $i'$, {\em etc.}\xspace) 
for scenarios, keywords and slots, respectively.

\item
Variables such as $m$, $n$, $\K_j$, $d_j$, $B$, $\eps_i$, $a_{i,j}$, $a_{i,j,k}$, $c_{i,j}$, $c_{i,j,k}$, $x_j$, $x_{j,k}$, $\mbox{payoff}$, $\mbox{payoff}_i$, $\kappa$, $s$ and 
$B$, when used in the context of the stochastic budget optimization models, 
will be used for their intended meanings as described in Sections~\ref{sec-singleslot}---\ref{sec-multislot}. 

\item
Note that:
\begin{itemize}
\item 
$m$, $n$, $d_j$, $B$, $a_{i,j}$, $a_{i,j,k}$, $c_{i,j}$, $c_{i,j,k}$ and $s$
are positive integers; 

\item
$0\leq\eps_i\leq 1$ and $\sum_{i=1}^m\eps_i=1$; 

\item
$1\leq\kappa=O\left(\poly(\log (m+n))\right)$ is an integer. We refer to this in the sequel by the phrase ``$\kappa$ is a small integer''.
\end{itemize}

\item
The size of an input instance of our \SBO\ problems, which we will denote by {\sf size-of-input} and which is {\em crucial 
in differentiating polynomial-time algorithms from pseudo-polynomial-time algorithms}, is as follows:
\begin{itemize}
\item 
For \ISSBO\ and \FSSBO: 
\[
\mbox{\sf size-of-input}=\poly\left(m+n+\left(\max_{\substack{1\leq i\leq m \\ 1\leq j\leq n}} \log_2 a_{i,j}\right)+\left(\max_{\substack{1\leq i\leq m \\ 1\leq j\leq n}}\log_2 c_{i,j}\right)+\left(\max_{1\leq i\leq m}\frac{1}{\eps_i}\right)\right).
\]

\item 
For \IMSSBO\ and \FMSSBO, 
\[
\mbox{\sf size-of-input}=\poly\left(s+m+n+\left(\max_{\substack{1\leq i\leq m \\ 1\leq j\leq n \\ 1\leq k\leq s}} \log_2 a_{i,j,k}\right)+\left(\max_{\substack{1\leq i\leq m \\ 1\leq j\leq n \\ 1\leq k\leq s}}\log_2 c_{i,j,k}\right)+\left(\max_{1\leq i\leq m}\frac{1}{\eps_i}\right)\right).
\]
\end{itemize}
\end{itemize}
On rare occasions, if we need to reuse the above-mentioned indices or variables and thus 
deviate from these conventions, the accompanying text will make the deviation clear.

\section{Summary of Results and Proof Techniques}
\label{sec-results}

~\cite{MPS07} left the computational complexity issues of the scenario model as the 
main open problem after showing that both the integral and fractional versions of this problem, even for single slot
case, are $\NP$-hard and noting that {\em no} non-trivial approximability results are known. While prior results for (S)BO 
problems exploit insights from the Knapsack problem to associate some potential payoff with each keyword, 
a central difficulty encountered in directly applying those techniques for our models 
is that payoff from a keyword can be very different from one scenario to another. 

\subsection{Summary of Results}

We provide a slightly coarse summary of the results obtained in this paper; precise bounds 
are available in the corresponding technical section that proves the result.

\subsection*{Main Results}

\begin{description}
\item[(R1) (Approximation algorithms):]
We provide algorithms that run in {\em near-linear} time and 
achieve the following approximation ratios\footnote{The reader is reminded that $\kappa=O\left(\poly(\log (m+n))\right)$.}: 
\begin{itemize}
\item
$\min\left\{O(m),O\!\left(\kappa\,\log d_n\right)\right\}$-approximation
for both \ISSBO\ and \FSSBO\ and, 

\item
$\min\left\{O(m),O\left(s\,\kappa\,\log \Delta\,\log^2(m+n)\right)\right\}$-approximation for 
\IMSSBO\ and \FMSSBO, where $\Delta=\max_{j,k} d_{j,k}$.
\end{itemize}

\item[(R2) (Approximation hardness for the single and multi slot cases)]
We show that, unless $\mathsf{ZPP}\!=\!\NP$, 
there exist instances of \ISSBO\ and \FSSBO, with $n$ keywords and $m=n$ scenarios each with equal probability,
such that {\em any} polynomial-time algorithm for solving these problems must have an approximation ratio 
of any one of the following (for any constant $0<\eps<1$): 
\begin{itemize}
\item 
$\Omega\left(m^{1-\eps}\right)$ (and, thus, also $\Omega\left(n^{1-\eps}\right)$), or

\item  
$\Omega\left(\kappa\,\log^{1-\eps} d_n\right)$.  
\end{itemize}
This {\em almost matches} the upper bounds in {\bf (R1)}. 
Thus, we cannot in general improve the approximation bound in {\bf (R1)}. 

Since \SSBO\ problems are special case of \MSSBO\ problems for $s=1$, 
the approximation hardness bounds for \SSBO\ can be extended to \MSSBO,
providing lower bounds of the form 
$\Omega\left(m^{1-\eps}\right)$, $\Omega\left(n^{1-\eps}\right)$, or
$\Omega\left(\log\kappa\cdot\log^{1-\eps} d_n\right)$
for \MSSBO\ instances with $n$ keywords, $m=n$ scenarios and $s$ slots.
We also show that \IMSSBO\ is {\sf MAX-SNP}-hard for $s=2$
even when $\kappa=1$ and $c_{j,k}=1$ for all $j$ and $k$. 
\end{description}

\subsection*{Other Results}

In addition to the main results, we 
also prove a number of other results dealing with variations and special
cases of our problems.
\begin{description}
\item[Fixed parameter tractability issues:]
For certain parameter ranges of practical interest 
we show that these optimization problems can be solved efficiently.
If $m$ or $ns$ is fixed,  \FMSSBO\ has a polynomial time 
solution with an absolute error of $\delta$ for any fixed $\delta>0$. 
If additionally bids are polynomial in size, \IMSSBO\ also has  
a polynomial time 
solution with an absolute error of $\delta$ for any fixed $\delta>0$.

\item[Limitations of semi-definite programming based approaches:]
The lower bounds in {\bf (R2)} 
have $\eps<1$ and thus leave a ``very small'' gap between this lower bound 
and the upper bounds described in {\bf (R1)}.
It is natural to ask if the gap could be eliminated; for example 
can we design an approximation algorithm for the special case for $\kappa=1$ 
whose approximation ratio is, say, $o\left(\frac{m}{\log m}\right)$ or $o\left(\frac{\log d_n}{\log\log d_n}\right)$?
Although we are unable to provide a concrete proof that such a polynomial time 
approximation algorithm does not exist, we nonetheless observe that the 
natural semidefinite programming relaxation will not work since it has a large
integrality gap of $\frac{m}{2}=\Theta\left(\frac{\log d_n}{\log\log d_n}\right)$.

\item[Dual of \SSBO\ problems:]
Finally, in some cases, the {\em dual} of the stochastic budget optimization problem may be of interest, where we are given a target
expected number of clicks and the goal is to minimize the expected budget spent while reaching the target. We present
some exact and approximate results for this dual version of the problem. 
\end{description}

\subsection{Brief Overview of Proof Techniques}

In general, budget optimization problems are akin to knapsack problems\footnote{See for example, \url{http://algo.research.googlepages.com/ec09_pub.pdf}}.
But the stochastic budget optimization problems studied in this paper are different because their budgets are ``soft'', \IE, they can be exceeded, if under a suitable scaling
they meet the budget constraint, and this improves the objective function. 
The stochastic budget optimization problems can be more insightfully thought of as special bipartite quadratic programs (these with $\pm 1$ variables 
correspond to Grothendieck's inequality with a nice history, but we have $0/1$ variables). Standard approaches to solving other special cases of 
quadratic programs, for example, using relaxations via semi-definite programming, do not provably work as we show. Instead, for upper bounds, we 
take alternative combinatorial approaches. 
For showing hardness results, we use intuitions from connections of our problems to these 
quadratic programs. 
For one proof, we show reduction from the hard instances of the 
{\em maximum independent set} problem~\cite{H99} on graphs 
to the bipartite $0/1$ quadratic integer programming reformulations of \FSSBO\ and \ISSBO. 
While anecdotally one may indeed believe these problems to be computationally hard, our results show that this is 
not true for many ranges of parameters of interest, but do identify the parameter settings that make them
computationally hard. Taken together, our results are the {\em first known} non-trivial complexity results for 
stochastic budget optimization problems under the scenario model beyond $\NP$-hardness. 

\section{\SBO\ Problems and Bipartite Quadratic Integer Programs} 
\label{whyqp}

In this section we show how to reformulate various \SBO\ problems as {\em bipartite quadratic integer programs} (\QIP).
These reformulations are heavily used in later proofs in the paper.
A bipartite quadratic program is a quadratic program in which there is a bipartition
of variables such that every term involves at most one variable from each
partition. A well-known example of such a (strict) quadratic program on variables 
taking $\pm 1$ values is the so-called Grothendieck's inequality~\cite{AN04}. However,
as will show later, our quadratic program differs significantly in nature from
this inequality.

\subsection{\SSBO\ and \QIP} 

\begin{figure}[htbp]
\begin{tabular}{m{3in}l}\toprule
{\bf (* Quadratic program {\bf (Q1)} *)} & 
   {\bf (* Quadratic program {\bf (Q2)} *)} \\
(* $w_{i,j}=y_{i,j}\,c_{i,j}$ for all $i$ and $j$ *) & 
   (* $w_{i,j,k}=y_{i,j,k}\, c_{i,j,k}$ for all $i$, $j$ and $k$ *) \\
{\sf maximize} $\displaystyle\sum_{i=1}^m \sum_{j=1}^n \alpha_i\, x_j\, y_{i,j}$ &
   {\sf maximize} $\displaystyle\sum_{i=1}^m \alpha_i\, \left(\sum_{j=1}^n \sum_{k=1}^s x_{j,k}\, y_{i,j,k}\right)$ \\
{\sf subject to} &
   {\sf subject to} \\
\hspace*{0.2in} $\displaystyle\forall\, 1\leq i\leq m\colon \,\,\alpha_i \left(\sum_{j=1}^n w_{i,j} x_j\right)\leq B_i$ &
   \hspace*{0.2in} $\displaystyle\forall\,1\leq i\leq m\colon\,\,\alpha_i\left(\sum_{j=1}^n\sum_{k=1}^s w_{i,j,k}\, x_{j,k}\right)\leq B_i$ \\
\hspace*{0.2in} $\forall\, 1\leq i\leq m\colon\,\,0\leq \alpha_i\leq 1$ &
   \hspace*{0.2in} $\forall\,1\leq j\leq n\colon\,\,\sum_{k=1}^s x_{j,k}\leq 1$ \\
\hspace*{0.2in} $\forall\,1\leq j\leq n\colon\,\,\, 0\leq x_j \leq 1$ & 
   \hspace*{0.2in} $\forall\,1\leq i\leq m\colon\,\, 0\leq \alpha_i\leq 1$ \\
   & 
   \hspace*{0.2in} $\forall\, 1\leq j\leq n\,\,\forall\,1\leq k\leq s\colon\,\,0\leq x_{j,k} \leq 1$ \\
\bottomrule
\end{tabular}
\caption{\label{fig1}Quadratic Integer Programs for {\sc Sbo} problems.
$Y$ is a matrix with non-negative entries ($y_{i,j}$ for {\bf (Q1)} and $y_{i,j,k}$ for {\bf (Q2)}) 
and $B_1,B_2,\ldots,B_m$ are positive real numbers.}
\end{figure}

We show how to reformulate \SSBO\ as a bipartite quadratic integer program.
Consider the quadratic program~{\bf (Q1)} in Fig.~\ref{fig1}.
By ``integral version'' of {\bf (Q1)} we refer to replacing the
constraint $0\leq x_i \leq 1$ by $x_i\in\{0,1\}$. 

\begin{Proposition}\label{PROP2}
The quadratic program {\bf (Q1)} and its integral version are equivalent to
\ISSBO\ or \FSSBO, respectively.
\end{Proposition}

\begin{proof}
Consider an instance of \SSBO.
Let $y_{i,j}=\eps_i\, a_{i,j}$, $w_{i,j}=c_{i,j}\, y_{i,j}$ and 
$B_i=\eps_i\,B$. 
Then, 
the inequality 
$\sum_{j=1}^n a_{i,j} c_{i,j} x_j \leq B$
becomes 
$\sum_{j=1}^n y_{i,j} c_{i,j} x_j \leq B_i
\equiv\sum_{j=1}^n w_{i,j} x_j\leq B_i$ 
and the fraction
$\dfrac{B}{\sum_{j=1}^n a_{i,j} c_{i,j} x_j}$
becomes
$\dfrac{B_i}{\sum_{j=1}^n w_{i,j}x_j}$.
Conversely, given an instance of {\bf (Q1)}, 
let 
$B=\sum_{i=1}^m B_i$, 
$\eps_i=\dfrac{B_i}{B}$
and 
$a_{i,j}=\dfrac{y_{i,j}}{\eps_i}$.
Thus,
$\eps_i\, a_{i,j}=y_{i,j}$,
the inequality
$\sum_{j=1}^n w_{i,j}x_j \leq B_i\equiv\sum_{j=1}^n y_{i,j}c_{i,j} x_j \leq B_i$
is the same as 
$\sum_{j=1}^n a_{i,j}c_{i,j} x_j \leq B$
and
the fraction
$\dfrac{B_i}{\sum_{j=1}^n w_{i,j}x_j}$
is the same as
$\dfrac{B}{\sum_{j=1}^n a_{i,j} c_{i,j} x_j}$.
Thus, in the sequel, we assume such a correspondence.

Now, consider a solution vector $\x=(x_1,x_2,\dots,x_n)$ and $\balpha=(\alpha_1,\alpha_2,\dots,\alpha_m)$ for {\bf (Q1)}.
Then $\x$ also defines a solution vector for 
\SSBO.
We must verify that this is indeed a valid solution vector with
a correct expected payoff.
Let $Q_i=\sum_{j=1}^n w_{i,j} x_j$.
If $\alpha_i Q_i < B_i$
then 
$\alpha_i=1$ since otherwise the solution for {\bf (Q1)} can be further improved, 
and then 
$\Ex[\mbox{payoff}_i]=\sum_{j=1}^n y_{i,j} x_{j}$, which is correct.
Otherwise 
$\alpha_i Q_i=B_i$
and 
then 
$\Ex[\mbox{payoff}_i]=\dfrac{B_i}{B_i/\alpha_i}\sum_{j=1}^n y_{i,j} x_{j}=
\alpha_i\sum_{j=1}^n y_{i,j} x_{j}$, 
which is also correct.
This shows that for every instance of {\bf (Q1)} there is a corresponding
instance of \SSBO\ with the same expected payoff.

Now, consider a solution vector $\x$ for \SSBO.
Then, if $Q_i\geq B_i$ 
then $\alpha_i=B_i/Q_i$ 
otherwise
$\alpha_i=1$.
It is easy to see in the same manner that this provides 
a valid solution of {\bf (Q1)} with the same objective value.
\end{proof}

\begin{wrapfigure}[8]{r}{2.6in}
\begin{tabular}{l}\toprule
{\sf maximize} $\alpha_i \sum_{j=p}^q y_{i,j}\,x_j$  \\
{\sf subject to} $\alpha_i \left(\sum_{j=p}^q c_{i,j}\,y_{i,j}\,x_j \right) \leq B_i$ \\
\hspace{0.2in} $\forall\,p\leq j\leq q\colon\,\,0\leq x_j\leq 1$ \\
\bottomrule
\end{tabular}
\caption{\label{fig111}{\sf LP} for $i$th row and $p$th through $q$th column of $Y$.}
\end{wrapfigure}

\paragraph{Relationship to the Standard Knapsack Problems} 
If $m=\kappa=1$ and $\alpha_1$ is set to a fixed constant, then 
{\bf (Q1)} reduces a special linear program which 
is equivalent to the so-called (fractional)
knapsack problem which is well-studied in the literature. 
Extending this analogy,
by the phrase ``{\sf the standard fractional knapsack problem corresponding to the $i$th row
of} and $p$th through $q$th column of $Y$'', we will mean 
the linear program as shown in Fig.~\ref{fig111} (it is easy to see that there is an optimal solution of 
this linear program in which 
$\alpha_i=\min\left\{1,\frac{B_i}{\left(\sum_{j=p}^q c_{i,j}\,y_{i,j}\right)}\right\}$).
Since $d_p\leq d_{p+1}\leq\dots\leq d_q$, the following well-known fact follows.

\begin{fact}{\rm\cite{GJ79}}\label{fact-knapsack} 
An optimal solution to the linear program in Fig.~\ref{fig111} 
(``optimal payoff for the $i$th row and $p$th through $q$th column of $Y$'') 
is a ``prefix solution'', \IE, there is an index $j\,\pmb{'}$ 
such that $x_j=1$ for $j<j\,\pmb{'}$, $0<x_{j\,\pmb{'}}\leq 1$ and $x_j=0$ for $j>j\,\pmb{'}$.
\end{fact}

\subsection{\MSSBO\ and QIP}

The quadratic programming reformulation of \MSSBO\
can also be obtained in a similar manner and is shown as~{\bf (Q2)} in Fig.~\ref{fig1}.

\section{Poly-logarithmic Approximations for \SSBO\ and \MSSBO\ (main result (R1))}
\label{sec-polylog}

\begin{Theorem}[Near-linear time approximation]\label{KNAPSACK}
There is a 
\begin{description}
\item[(i)]
$\min\left\{O(m),O\left(\kappa\log d_n\right)\right\}$-approximation for 
both \ISSBO\ and \FSSBO; 

\item[(ii)]
$\min\left\{O(m),O\left(s\kappa\log \Delta\right)\right\}$-approximation for 
\FMSSBO\ and 

\item[(iii)]
$\min\left\{O(m),O\left(s\kappa\log\Delta\log^2(m+n)\right)\right\}$-approximation for 
\IMSSBO
\end{description}
where, for {\bf (ii)} and {\bf (iii)}, $\Delta=\max_{j,k} d_{j,k}$.
All these algorithms can be implemented in linear or near-linear time using 
standard data structures and algorithmic techniques.
\end{Theorem}

In the rest of this section, we prove the above theorem. 
As a first attempt, one might be tempted to use recent techniques in designing efficient
algorithms for multiple-knapsack problems~\cite{CK00,J09} for our problem; however it is not 
difficult to design examples where such approaches fail badly since 
our budget constraints are ``soft'' (they can be exceeded if scaling them gives better payoff) and our probabilities are ``arbitrary''. As a second
attempt, one might take our quadratic programming reformulation as discussed in 
Section~\ref{whyqp} and semidefinite-programming based rounding approach such as in~\cite{GW95}. However, 
it can be shown that 
the integrality gap of such a reformulation is very large.
The failure of these natural approaches shows the difficulty of the problems. 
Thus, we are led to explore other combinatorial approaches to provide the desired approximation.

\subsection{$O(m)$-approximation for \ISSBO\ and \FSSBO}
\label{simple}
 
To get a $O(m)$-approximation we can do the following.
For each $i$ we solve the standard (integer or fractional)
knapsack problem for the $i$th row of $Y$; let $p_i$ be the
value of an optimal solution.
Then, take the best of these solutions, say of value $p=\max_{1\leq i\leq m}\{p_i\}$.
Each fractional knapsack problem can be solved exactly in $O(n\log n)$ time~\cite{GJ79} and a $O(n\log n)$ time greedy $2$-approximation
algorithm for the integer knapsack problem is also well known~\cite{L79}.

We now note that $\Ex[\mbox{payoff}]\leq\sum_{i=1}^m p_i$. 
Indeed, consider an optimal solution of \SSBO.
If $\alpha_i=1$, then by definition of $p_i$ we 
have $\Ex[\mbox{payoff}_i]\leq p_i$. 
If $\alpha_i<1$, then we set $\alpha_i=1$ and set 
a new value of $x_j$ as $x_j'=\alpha_ix_j$. This does
not change $\Ex[\mbox{payoff}_i]$ and now we again have 
$\Ex[\mbox{payoff}_i]\leq p_i$. 
Thus, we have $p\geq \Ex[\mbox{payoff}]/m$. 

If $p=p_i$ for some $i$, then the solution of the knapsack problem
of value $p$ can be extended to a solution of \SSBO\ by 
setting $\alpha_{i'}=0$ for $i'\neq i$.

\subsection{$O\left(\kappa\log d_n\right)$-approximation for \ISSBO\ and \FSSBO}

\subsubsection*{Case of $\mathbf{\kappa=1}$: Uniform Cost Model}

\begin{figure}
\begin{flushleft}
\begin{tabular}{lp{6in}}\toprule
{\bf 1.} & Partition the keywords into {\em maximal} groups such that if 
a group $G$ contains $p$th through \newline 
\hspace*{0.2in} $q$th keyword then $d_q/d_p\leq 2$ and $d_{q+1}/d_p>2$.  
\newline 
Let $\G$ be the set of such groups. \\
{\bf 2.} & {\bf For} each group $G\in\G$ consisting of keywords, say $\K_p,\K_{p+1},\ldots,\K_q$, {\bf do} \newline
           \mbox{ }\mbox{ } Set $x_j=1$ for every $p\leq j\leq q$ and set $x_j=0$ for all other $j$; \newline 
           \mbox{ }\mbox{ } let $\Ex[{\rm payoff}']$ be the payoff of this solution \\
{\bf 3.} & Output the best of the solutions obtained in {\bf 2}. \\
\bottomrule
\end{tabular}
\caption{\label{grouping}Algorithm for the case of $\kappa=1$.} 
\end{flushleft}
\end{figure}

The algorithm is shown in Fig.~\ref{grouping}.
Consider a group $G\in\G$ consisting of the keywords 
$\K_p,\K_{p+1},\ldots,\K_q$.
By the ``\SSBO\ problem on $G$'' we mean the instance of the \SSBO\ problem
in which our click input consists of the submatrix 
$
Y_{p,q}=
\begin{pmatrix}
y_{1,p} & y_{1,p+1} & \ldots & y_{1,q} \\
\vdots  & \vdots    & \ldots & \vdots \\
y_{m,p} & y_{m,p+1} & \ldots & y_{m,q} \\
\end{pmatrix}
$
of $Y$ 
containing all rows and $p$th through $q$th columns,
the costs-per-click $d_p,\ldots,d_q$, the budgets $B_1,\ldots,B_m$, 
and the selection variables $x_p,\ldots,x_q$.
Let 
$\Ex[\mbox{payoff}_G]$ 
be the value of expected payoff of an optimal solution for this subproblem.
Since $\max_{G\in\G}\Ex[\mbox{payoff}_G]\geq\frac{\Epayoff}{|\G|}$ and $|\G|=O(\log d_n)$,
the following lemma proves the desired approximation bound.

\begin{Lemma}\label{PAYOFF-LEMMA}
$\Ex[{\rm payoff}']\geq\dfrac{\Ex[{\rm payoff}_{G}]}{2}$. 
\end{Lemma}

\begin{proof}
We only need to prove the lemma for the case when $\Ex[{\rm payoff}_G]$ is the total expected payoff of an optimal solution of the \FSSBO\ problem on $G$ since obviously
the total expected payoff of an optimal solution of the \ISSBO\ problem on $G$ is no more than $\Ex[{\rm payoff}_G]$.
Let $D=\sum_{j=p}^q d_j\,y_{i,j}$ and $\beta=|G|$. 
By our choice of the group $G$, 
\[
d_p\sum_{j=p}^q y_{i,j}\leq D\leq d_q\sum_{j=p}^q y_{i,j}\leq 2\,d_p\sum_{j=p}^q y_{i,j}.
\]
Using the quadratic programming formulation {\bf (Q1)} and remembering that $c_{i,j}=d_j$ when $\kappa=1$, 
the \FSSBO\ instance on $G$ is equivalent to the following quadratic program {\bf (Q3)}:

\begin{center}
\begin{tabular}{l}\toprule
{\bf (* Quadratic program (Q3) *)} \\
maximize $\sum_{i=1}^m \alpha_i \left(\sum_{j=p}^q y_{i,j}\,x_j \right)$ \\
subject to $\forall\, 1\leq i\leq m\colon \,\,\alpha_i \left(\sum_{j=p}^q d_j\,y_{i,j}\,x_j\right)\leq B_i$ \\
\hspace*{0.7in} $\forall\, 1\leq i\leq m\colon\,\,0\leq \alpha_i\leq 1$ \\
\hspace*{0.7in} $\forall\,p\leq j\leq q\colon\,\,\, 0\leq x_j \leq 1$ \\
\bottomrule
\end{tabular}
\end{center}

\noindent
Fix any optimal solution for our \FSSBO\ instance on $G$, \IE, fix an optimal solution vector 
$(\alpha_1^\ast,\alpha_2^\ast,\dots,\alpha_m^\ast)$ and $(x_p^\ast,x_{p+1}^\ast,\dots,x_q^\ast)$ of {\bf (Q3)}.
In our solution sets $x_p=x_{p+1}=\cdots=x_q=1$; thus $\alpha_i=\min\left\{1,\,\frac{B_i}{D}\right\}$ for every $i$
and $x_j\geq x_j^\ast$ for every $p\leq j\leq q$. 
\begin{description}
\item[Case 1: $D\leq B_i$]. 
Then, $\alpha_i=1\geq\alpha_i^\ast$, $x_j=1\geq x_j^\ast$ for $p\leq j\leq q$,
and thus 
\[
\alpha_i\left(\sum_{j=p}^qy_{i,j}\,x_j\right)\geq\alpha_i^\ast\left(\sum_{j=p}^q y_{i,j}\,x_j^\ast\right).
\]

\item[Case 2: $D>B_i$]. 
Then, $\alpha_i=\frac{B_i}{D}$.
Now, we have 
\begin{gather*}
\alpha_i \left(\sum_{j=p}^q y_{i,j}\,x_j\right)
=\left(\frac{B_i}{D}\right)\times\sum_{j=p}^q y_{i,j}
\geq
\left(\frac{B_i}{D}\right)\times\left(\frac{\sum_{j=p}^q d_j\,y_{i,j}}{d_q}\right)
=
\frac{B_i}{d_q}
\geq
\frac{1}{2}\times\frac{B_i}{d_p}
\\
\alpha_i^\ast \left(\sum_{j=p}^q y_{i,j} x_j^\ast\right) \leq \frac{B_i}{\sum_{j=p}^q y_{i,j} d_j x_j^\ast}\times \left(\sum_{j=p}^q y_{i,j} x_j^\ast\right) \leq \frac{B_i}{d_p}
\end{gather*}
where the inequality for $\alpha_i^\ast$ comes directly from the constraints of {\bf (Q3)}.
\end{description}
Thus, combining both cases, we have 
\[
\Ex[{\rm payoff}']
=
\sum_{i=1}^m \alpha_i \left(\sum_{j=p}^q y_{i,j}\,x_j\right)
\geq 
\frac{1}{2}\times\sum_{i=1}^m \alpha_i^\ast\left(\sum_{j=p}^q y_{i,j}\,x_j^\ast\right)
=
\frac{\Ex[{\rm payoff}_{G}]}{2} 
\]
\end{proof}

\subsubsection*{Case of $\mathbf{\kappa>1}$: General Single-slot Model}

Using our $\delta$-approximation algorithm for \USSBO\ (for $\delta=O\left(\log d_n\right)$) 
as outlined in Fig.~\ref{grouping}, we show how to use it as a subroutine to get a 
$\kappa\,\delta=O\left(\kappa\log d_n\right)$-approximation for 
\ISSBO\ (and, hence, also for \FSSBO). The algorithm is shown in Fig.~\ref{simple-scaling}. 

\begin{figure}[h]
\begin{tabular}{lp{6in}}\toprule
{\bf 1.} & Replace (truncate) each $c_{i,j}$ by its {\em new} value $c_{i,j}'=d_j$. \\
{\bf 2.} & Use the approximation algorithm in Fig.~\ref{grouping} with these new truncated values of $c_{i,j}$'s. \\ 
         & Let $\x=(x_1,x_2,\dots,x_n)$ and $\balpha=(\alpha_1,\alpha_2,\dots,\alpha_m)$ be the solution vectors returned. \\
{\bf 3.} & Output $\x$ and $\balpha'=(\alpha_1',\alpha_2',\dots,\alpha_m')=\left(\frac{\alpha_1}{\kappa},\frac{\alpha_2}{\kappa},\dots,\frac{\alpha_m}{\kappa}\right)$ as our solution. \\
\\
\bottomrule
\end{tabular}
\caption{\label{simple-scaling}$O(\kappa\log d_n)$-approximation algorithm for \ISSBO.} 
\end{figure}

\noindent
We use the following notations:
\begin{itemize}
\item
$\x^\ast=(x_1^\ast,x_2^\ast,\dots,x_n^\ast)$ and $\balpha^\ast=(\alpha_1^\ast,\alpha_2^\ast,\dots,\alpha_m^\ast)$ are the solution vectors  
for an optimal solution of our (original) instance of \SSBO, and 
$\Ex[\mbox{payoff}^\ast]=\sum_{i=1}^m \alpha_i^\ast \left(\sum_{j=1}^n y_{i,j}x_j^\ast\right)$ 
is the total expected payoff of this optimal solution.

\item 
$\x^+=(x_1^+,x_2^+,\dots,x_n^+)$ and $\balpha^+=(\alpha_1^+,\alpha_2^+,\dots,\alpha_m^+)$ are the solution vectors  
for an optimal solution of the truncated instance of \SSBO, and 
$\Ex[\mbox{payoff}^+]=\sum_{i=1}^m \alpha_i^+ \left(\sum_{j=1}^n y_{i,j}x_j^+\right)$ 
is the total expected payoff of this optimal solution.

\item
$\Ex[\mbox{payoff}]=\sum_{i=1}^m \alpha_i' \left(\sum_{j=1}^n y_{i,j}x_j\right)$ 
is the total expected payoff of the solution obtained by using the algorithm in Fig.~\ref{simple-scaling}.
\end{itemize}

\begin{Proposition}\label{ROUNDING}
The following statements are true:
\begin{description}
\item[(a)]
$\x$ and $\balpha'$ correspond to a valid solution of the \SSBO\ instance.

\item[(b)]
$\Ex[{\rm payoff}^+]\geq\Ex[{\rm payoff}^\ast]$. 

\item[(c)]
$\Ex[{\rm payoff}]\geq\frac{\Ex[{\rm payoff}^+]}{\kappa}$. 
\end{description}
Thus the algorithm in Fig.~\ref{simple-scaling} is a $O\left(\kappa\log d_n\right)$-approximation.
\end{Proposition}

\begin{proof}~\\
\noindent
{\bf (a)} 
$\alpha_i'c_{i,j}\!=\!\frac{\alpha_i}{\kappa}c_{i,j}\leq\frac{\alpha_i}{\kappa}\,\kappa\, d_j\!=\!\alpha_i d_j$,
thus 
$\displaystyle\alpha_i\left(\sum_{j=1}^n y_{i,j}\,d_j\,x_j\right)\leq B_i$
implies 
$\displaystyle\alpha_i'\left(\sum_{j=1}^n y_{i,j}\,c_{i,j}\,x_j\right)\leq B_i$.

\noindent
{\bf (b)} 
The solution vectors $\x^\ast$ and $\balpha^\ast$ for an optimal solution of the \SSBO\ instance 
is also a valid (not necessarily optimal) solution vector for the truncated instance of \SSBO\ since 
$c_{i,j}'\leq c_{i,j}$.

\vspace*{0.1in}
\noindent
{\bf (c)} 
This follows since $\alpha_i'=\frac{\alpha_i}{\kappa}$.
\end{proof}

\subsection{Approximation Bounds for \FMSSBO\ and \IMSSBO}

\begin{wrapfigure}{r}{0pt}
\begin{tabular}{l}\toprule
{\bf (* Quadratic program (Q4) *)} \\
{\sf maximize} $\alpha_i\left(\sum_{j=1}^n \sum_{k=1}^s y_{i,j,k}\,x_{j,k}\right)$ \\
{\sf subject to} \\
\hspace*{0.2in} $\alpha_i\left(\sum_{j=1}^n\sum_{k=1}^s w_{i,j,k}\, x_{j,k}\right)\leq B_i$ \\
\hspace*{0.2in} $\forall\,1\leq j\leq n\colon\,\,\sum_{k=1}^s x_{j,k}\leq 1$ \\
\hspace*{1.2in} $0\leq \alpha_i\leq 1$ \\
\hspace*{0.2in} $\forall\, 1\leq j\leq n\,\,\forall\,1\leq k\leq s\colon\,\,0\leq x_{j,k} \leq 1$ \\
\bottomrule
\end{tabular}
\caption{\label{mkcp}\MSSBO\ restricted to the $i$th scenario.}
\end{wrapfigure}

To get a $O(m)$-approximation we follow the same approach as in Section~\ref{simple}. 
For each $i$ we solve the restriction of the \MSSBO\ problem on the $i$th scenario, \IE, 
the quadratic program {\bf (Q4)} as shown in Fig.~\ref{mkcp},
and then take the best of these solutions.
It is easy to see that an optimal solution of {\bf (Q4)} satisfies 
$\alpha_i=\min\left\{1,\frac{B_i}{\sum_{j=1}^n\sum_{k=1}^s w_{i,j,k}x_{j,k}}\right\}$.
For any fixed value of $\alpha_i$, {\bf (Q4)} is known in the literature as the {\em multiple-choice} Knapsack
problem with $s\,n$ objects divided into $n$ classes and a knapsack capacity of $B_i/\alpha_i$; 
a $O(1)$-approximation algorithm for this problem that runs in $O\left(ns^2\right)$ time 
is known~\cite{L79}. 

We next show that algorithms for the single-slot case can be used for the multi-slot model
with appropriate multiplicative factors in the approximation ratio.

\begin{Lemma}\label{sfactor}
There exists a $O(s\,\kappa\,\log\Delta)$-approximation (respectively, 
$O\left(s\,\log^2(m+n)\,\kappa\,\log\Delta\right)$-approximation) algorithm 
for \FMSSBO\ (respectively, \IMSSBO).
\end{Lemma}

\begin{proof}
We first prove our claim for \FMSSBO.
Consider the quadratic program {\bf (Q2)'} obtained from the quadratic program {\bf (Q2)} for 
\FMSSBO\ by removing 
the constraints 
$\sum_{k=1}^s x_{j,k}\leq 1$ for $1\leq j\leq n$.
If $\mathsf{OPT}$ and $\mathsf{OPT}'$ are the optimal values of the objective functions of
{\bf (Q2)} and {\bf (Q2)'}, respectively, then obviously 
$\mathsf{OPT}'\geq\mathsf{OPT}$.
A straightforward inspection shows that {\bf (Q2)'} can be written down in the same form as {\bf (Q1)} 
with $s\,n$ variables and $m$ constraints.
Thus, using the already proven result of Theorem~\ref{KNAPSACK}{\bf (i)} 
we obtain a solution for {\bf (Q2)'}
whose objective value is 
$
\frac{\mathsf{OPT}'}{\kappa\log(\max_{j,k}d_{j,k})}
=
\frac{\mathsf{OPT}'}{\kappa\log\Delta}
\geq 
\frac{\mathsf{OPT}}{\kappa\log\Delta}
$
To convert this to a solution of \FMSSBO (\IE, to satisfy the constraints 
$\sum_{k=1}^s x_{j,k}\leq 1$ for each $j$) 
we divide each $x_{j,k}$ by $\sum_{k=1}^s x_{j,k}$
which decreases the total payoff by no more than a factor of 
$s$.

The result for \IMSSBO\ follows by translating the above worst-case approximation bound for \FMSSBO\ 
to a worst-case approximation of \IMSSBO\ via the following lemma.

\begin{Lemma}
\label{translate}
({\sc Approximating} \IMSSBO\ {\sc via} \FMSSBO)
Suppose that we have a $\eta$-approximation for \FMSSBO. Then, we also
have a $O(\eta\,\gamma)$ approximation for \IMSSBO\ where
$
\gamma=\left\{
\begin{array}{ll}
\log m, & \mbox{if $s=1$} \\
\log^2 (m+n), & \mbox{otherwise} \\
\end{array}
\right.
$
\end{Lemma}

\begin{proof}
For a particular value of the vector 
$\balpha=(\alpha_1,\alpha_2,\dots,\alpha_m)$,
{\bf (Q2)} reduces to a linear program on 
the variables $\x=(x_1,x_2,\dots,x_n)$. 
For ease of description, we consider the case of $s=1$ first (\IE, the case
of \FSSBO).
An inspection of {\bf (Q1)} reveals that
this linear program has exactly $n$ variables and 
$m$ inequalities, where the $ith$ inequality $D_i$ (for $1\leq i\leq m$) is of the form:
\[
D_i\,\,\pmb{\stackrel{\mathrm{def}}{\equiv}}\,\,\alpha_i\left(\sum_{j=1}^n w_{i,j}\,x_j\right)\leq B_i
\]
Consider a solution 
$\x^f=(x_1^f,x_2^f,\dots,x_n^f)$. 
and 
$\balpha^f=(\alpha_1^f,\alpha_2^f,\dots,\alpha_m^f)$,
of \FSSBO\ 
with $\cL=\sum_{i=1}^m \sum_{j=1}^n \alpha_i^fy_{i,j}\,x_j^f$ as the value of its objective. 
We may assume that $\cL>100\ln m$ since otherwise the approximation guarantee can be trivially achieved.
We employ the following randomized rounding scheme to transform this solution to a solution of \ISSBO:
\begin{itemize}
\item 
For $i=1,2,\dots,n$, we round $x_i^f$ randomly to $0$ and $1$ with 
probabilities $x_i^f$ and $1-x_i^f$, respectively. Let $x_i\in\{0,1\}$ be the resulting random variable.

\item 
We return $\x=(x_1,x_2,\dots,x_n)$ and $\balpha=(\alpha_1,\alpha_2,\dots,\alpha_m)$ 
as our solution where $\alpha_i=\frac{\alpha_i^f}{100\ln m}$ for $1\leq i\leq m$. 
\end{itemize}
Let $\cL'=\sum_{i=1}^m \sum_{j=1}^n \alpha_iy_{i,j}\,x_j$ 
be the new value of the objective
and let $\cE_i$ be the event that inequality $D_i$ holds for this randomized solution.
By linearity of expectation $\Ex[\cL']=\frac{\cL}{100\,\ln m}$. 
Consider the inequality $D_i$, and let $\alpha_i'=\frac{\alpha_i}{B_i+1}$.
By linearity of expectation, 
\[
\Ex\left[\alpha_i'\left(\sum_{j=1}^n w_{i,j} x_j\right)\right]
=
\frac{1}{100\ln m}\times\frac{1}{B_i+1}\times\alpha_i^f\left(\sum_{j=1}^n w_{i,j} x_j^f\right)
<
\frac{1}{100\ln m}\times\frac{B_i}{B_i+1}
\]
Since $\alpha_i'=\frac{\alpha_i}{B_i+1}$, 
$0\leq\alpha_i'w_{i,j}x_j=\frac{\alpha_i w_{i,j} x_j}{B_i+1}\leq \frac{B_i}{B_i+1}<1$
and thus 
$\alpha_i'w_{i,j}x_j$ can be thought of as an independent Poisson trial whose probability of success (a value of $1$)
is $\alpha_i'w_{i,j}x_j$ and probability of failure (a value of $0$) is $1-\alpha_i'w_{i,j}x_j$.
Thus, using standard Chernoff bound~\cite[Excercise 4.1]{MR95}, we get: 
\[
\Pr[\mbox{$\cE_i$ does not hold}]
=
\Pr\left[\alpha_i\left(\sum_{j=1}^n w_{i,j} x_j\right) > B_i\right]
=
\Pr\left[\alpha_i'\left(\sum_{j=1}^n w_{i,j} x_j\right)>\frac{B_i}{B_i+1}\right]
<
\bee^{-3\ln m}<\frac{1}{m^2}
\]
In a similar manner, one can show that 
$\Pr\left[\cL'<\frac{\cL}{200\ln m}\right]<\frac{1}{m}$. Thus, finally, using union bounds, we get
\[
\Pr\left[\cL'\geq\frac{\cL}{200\ln m}\bigwedge\left(\wedge_{i=1}^m \mbox{$\cE_i$ holds}\right)\right]
\geq
1-\Pr\left[\cL'<\frac{\cL}{200\ln m}\right]
-\left(\sum_{i=1}^m \Pr\left[\mbox{$\cE_i$ does not hold}\right]\right)
>
1-\frac{2}{m}
\]
Thus, we achieve the desired approximation bound with $1-o(1)$ probability.

For the case of $s>1$ (\IE, \FMSSBO), the same approach with some modifications works.
In a nutshell, we have $n$ additional constraints $F_j$ (for $j=1,2,\dots,n$) of
the form $\sum_{k=1}^s x_{j,k}\leq 1$.
Thus, the total number of inequalities/equalities is $m+n$ and we need to do
the analysis with ``$\ln (n+m)$'' replacing ``$\ln m$''. 
The only additional part that needs to be done is to show how to handle the $F_j$ constraints.
Notice that the set of variables involved in $F_j$ are disjoint from the set of variables 
in any other $F_{j'}$ for $j'\neq j$. 
After rounding, we have 
$\sum_{k=1}^s x_{j,k}\leq 100\ln (m+n)$.
We now select one of these variables $x_{j_1}$ to $x_{j,s}$, say $x_{j,\ell}$, 
such that
$x_{j,\ell}= \max_{1\leq k\leq s} \{\sum_{i=1}^m \alpha_i x_{j,k} y_{i,j,k}\,\}$, 
set $x_{j,\ell}=1$ and set $x_{j,k}=0$ for $k\neq\ell$. 
After all these normalizations, we loose an additional factor of $100\ln (m+n)$ 
and all constraints are satisfied.
\end{proof}

Note that the claim in Lemma~\ref{translate} is ``pessimistic'' 
in nature; indeed, as our claim in Theorem~\ref{KNAPSACK} shows, for arbitrary 
parameter range both \ISSBO\ and \FSSBO\ can be approximated to within the
same ratio. 
\end{proof}

\section{Approximation-hardness Results for \SSBO\ and \MSSBO\ (main result (R2))}
\label{hardnes} 

\newcommand{\gind}{\Delta_{\rm ind}}
\newcommand{\gq}{\Delta_{\rm Q1}}
\newcommand{\MIS}{{\sf MIS}}

\subsection{Approximation-hardness Bounds for \SSBO}

\begin{Theorem}[Logarithmic inapproximability]\label{LB}
There exist instances of \ISSBO\ and \FSSBO, with $n$ keywords and $m=n$ scenarios each with equal probability,
such that, unless $\mathsf{ZPP}\!=\!\NP$, {\em any} polynomial-time algorithm for solving these problems must have an approximation ratio 
of any one of the following:
\begin{itemize}
\item 
$\Omega\left(m^{1-\eps}\right)$ (and, thus, also $\Omega\left(n^{1-\eps}\right)$), or

\item  
$\Omega\left(\kappa\,\log^{1-\eps} d_n\right)$.  
\end{itemize}
where $0<\eps<1$ is any constant.
\end{Theorem}

\begin{proof}
We construct instances of \SSBO\ 
with $n$ keywords and $m=n$ scenarios such that, for\footnote{Remember that 
in Section~\ref{sec-notations} we fixed bounds on $\kappa$, namely, $\kappa=O\left(\poly(\log(m+n))\right)$.}
{\em any} $\kappa$ and {\em any} values of $c_{i,j}$ in the
range $[d_j,\kappa\,d_j)$, the claimed lower bound holds.
We use the reformulation of \FSSBO\ and \ISSBO\ as a bipartite
quadratic program {\bf (Q2)} as discussed in Section~\ref{whyqp}.

The standard maximum independent set (\MIS) problem is defined as follows. 
We are given an undirected graph $G=(V,E)$. A subset of vertices $V'\subseteq V$
is called {\em independent} if for {\em every} two vertices $u,v\in V'$ we have $\{u,v\}\not\in E$.
The goal is to find an independent subset of vertices of {\em maximum} cardinality.
It is known that \MIS\ cannot be approximated to within a factor of 
$|V|^{1-\eps}$ for any constant $0<\eps<1$ unless {\sf ZPP}$=\NP$~\cite{H99}. 

For notational simplicity, let $n=|V|$ and $a=n^{12}$. Set $m=n$.
Select an arbitrary order $v_1,v_2,\dots,v_n$ 
of the vertices in $V$. Intuitively, the $i$th column and the $(n+1-i)$th row
of $Y$ correspond to the vertex $v_i$ and the entries of the matrix $Y$ are such that they are 
$0$ above the reverse diagonal and encodes the adjacency of vertices of $G$ on or
below the reverse diagonal. Formally, 
\[
y_{i,j}=\left\{
\begin{array}{ll}
0 & \mbox{if $i+j<n+1$} \\
1 & \mbox{if $i+j=n+1$} \\
1 & \mbox{if $i+j>n+1$ and $\{v_{n-i+1},v_j\}\in E$} \\
0 & \mbox{if $i+j>n+1$ and $\{v_{n-i+1},v_j\}\not\in E$} \\
\end{array}
\right.
\]  
Fix $d_1,d_2,\dots,d_n$ as $d_1=1$ and $d_i=a\,d_{i-1}$ for $1<i\leq n$.
Thus, for all sufficiently large $n$,
$\frac{c_{i_1,j_1}}{c_{i_2,j_2}}\geq\frac{\frac{d_{j_1}}{\kappa}}{\kappa\,d_{j_2}}>n^6$ if $j_1>j_2$.
Let $B_i=c_{i,n+1-i}$ for $1\leq i\leq m=n$. Remembering that 
$w_{i,j}=c_{i,j}\,y_{i,j}$ for all $i$ and $j$, we have: 
\[
w_{i,j}=\left\{
\begin{array}{ll}
0 & \mbox{if $i+j<n+1$} \\
  & \mbox{or if $i+j>n+1$ and $\{v_{n-i+1},v_j\}\not\in E$} \\
c_{i,j} & \mbox{if $i+j=n+1$} \\
        & \mbox{or if $i+j>n+1$ and $\{v_{n-i+1},v_j\}\in E$} \\
\end{array}
\right.
\]
Note that $n^{1-\eps}=m^{1-\eps}=\Omega\left(\kappa\,\log^{1-\eps'} d_n\right)$, where $0<\eps'<1$ is a constant that depends on $\eps$,  
since $d_n=n^{12\,n}$ and $\kappa=\poly\left(\log(m+n)\right)=\poly\left(\log(n)\right)$.
Let $\gind$ and $\gq$ be the maximum number of independent vertices in $G$ and
an optimal value of the objective of the fractional or integral version of {\bf (Q1)}, respectively.

\begin{Lemma}\label{GAWK}
$\gq\geq\gind$.
\end{Lemma}

\begin{proof}
Consider an optimal solution $V'$ of MIS on $G$ with $|V'|=\gind$. 
We generate a solution of {\bf (Q2)} by setting 
\[
x_i=\alpha_{n-i+1}=\left\{
\begin{array}{ll}
1, & \mbox{if $v_i\in V'$} \\
0, & \mbox{otherwise} \\
\end{array}
\right.
\]
Note that, since $V'$ is an independent set,  
if $i+j>n+1$, $v_i\in V'$ and $\{v_i,v_j\}\in E$
then $v_j\not\in V'$ and thus $x_i=\alpha_{n-i+1}=1$ and $x_j=\alpha_{n-j+1}=0$.

First, we show that this is indeed a valid solution of {\bf (Q1)}. 
For any $1\leq i\leq n-1$, consider the constraint
\[
\alpha_{n-i+1}\left(\sum_{j=1}^n w_{n-i+1,j}\,x_j\right)\leq B_{n-i+1}.
\]
If 
$\alpha_{n-i+1}=0$, then
the constraint is obviously satisfied since $B_{n-i+1}>0$.
Otherwise, 
$\alpha_{n-i+1}=x_i=1$ and thus,
\[
\alpha_{n-i+1}\left(\sum_{j=1}^n w_{n-i+1,j}\,x_j\right)
=
\sum_{j=1}^n w_{n-i+1,j}\,x_j 
=
c_i+ \sum_{\substack{i+j>n+1 \\ \{v_i, v_j\}\in E}} w_{n-i+1,j}\,x_j
=
c_{n+1-i,i}
=B_{n+1-i}
\]
Thus, all the constraints {\em are} satisfied. Finally, the value of the objective
function is
\[
\sum_{i=1}^m \sum_{j=1}^n \alpha_i\,x_j\,y_{i,j}
=
\sum_{\substack{i+j=n+1 \\ v_j\in V'}} \alpha_i x_j
=
\sum_{v_j\in V'} x_j
=
\gind
\]
and thus $\gq\geq\gind$.
\end{proof}

For the other direction, we first need a normalization lemma.

\begin{Lemma}[Normalization lemma]\label{NORM}
Consider an optimal solution of {\bf (Q1)} with an objective value of $\gq$. 
Then, we can transform this solution to another solution of {\bf (Q1)} of objective value
$\gq'$ such that:
\begin{description}
\item[(a)]
$x_i\in\{0,1\}$ for each $i$; 

\item[(b)]
$\gq'\geq\gq -1$; and 

\item[(c)]
if $\{x_i,x_j\}\in E$ then $x_i+x_j\leq 1$.
\end{description}
\end{Lemma}

\begin{proof}
Suppose that we are given an optimal solution of {\bf (Q1)} with an objective value of 
$\gq$. First, we note some properties of this solution.

\begin{Proposition}\label{j1}
The following statements are true:
\begin{description}
\item[(i)]
for every $i$, $\alpha_{n-i+1} x_i\leq 1$, and 

\item[(ii)]
for every $i$ and $j$, if $i+j>n+1$ and $\{v_i,v_j\}\in E$ then $\alpha_j x_i\leq n^{-6}$. 
\end{description}
\end{Proposition}

\begin{proof}
Consider the constraint $\alpha_{n-i+1}\left(\sum_{j=1}^n w_{n-i+1,j} x_j\right)\leq B_{n-i+1}=c_{n-i+1,i}$.

Since $w_{n-i+1,i}=c_{n-i+1,i}$, {\bf (i)} follows.

{\bf (ii)} is equivalent to the claim that
$\alpha_{n-i+1}x_j\leq n^{-6}$ if $j>i$.
Since 
$\frac{c_{p,j}}{c_{q,i}}>n^6$ if $j>i$ (for any $p$ and $q$),
{\bf (ii)} follows.
\end{proof}

Now we show how to ``normalize'' this solution such that each variable 
$x_i$ is $0$ or $1$, and the total objective value does not decrease too much.
Let $\Gamma=\sum_{i+j\neq n+1}\alpha_i x_j y_{i,j}$.
By Proposition~\ref{j1}{\bf (ii)}, 
$\Gamma\leq n^2\times n^{-6}=n^{-4}$. 
Thus, setting 
$\Phi=\sum_{i+j=n+1}\alpha_i x_j y_{i,j}$,
it follows that 
$\Phi\leq\gq\leq\Phi+n^{-4}$.
Thus, subsequently we concentrate on the quantity $\Phi$. 

If $\alpha_{n-i+1}=0$ for some $i$, then we can 
set $x_i=0$ without changing the value of $\Phi$. 
Let $I=\{n-i+1\,|\,\alpha_{n-i+1}>0\mbox{ and } x_i>0\}$.
Consider the largest index $n-i+1\in I$.
There are two cases to consider:
\begin{description}
\item[Case 1: $x_i>n^{-3}$.]
By Proposition~\ref{j1}{\bf (i)},
$\alpha_{n-i+1}<n^{-3}$
and  
$\alpha_{n-j+1}x_j\leq\alpha_{n-j+1}<n^{-3}$ 
for every $j>i$ such that $\{v_i,v_j\}\in E$.

We set $\alpha_{n-i+1}=x_i=1$ and set $x_j=\alpha_{n-j+1}=0$ 
for every $j>i$ such that $\{v_i,v_j\}\in E$.
The change in $\Phi$ is at most 
$n\times n^{-3}=n^{-2}$.

\item[Case 2: $x_i\leq n^{-3}$.]
We set $\alpha_{n-i+1}=x_i=0$.
The change in $\Phi$ is at most 
$n^{-3}$.
\end{description}
We now remove the index $n-i+1$ from $I$ and continue with the next largest index.
We continue until $I=\emptyset$. Since $|I|\leq n$, the total change
in $\Phi$ is at most $n^{-1}<1-n^{-4}$.

To complete the proof, we select vertices $v_j$ in the independent 
set if $x_j=1$. 
\end{proof}

To finish the proof of Theorem~\ref{LB}, 
we simply select those vertices $v_i$ for the independent 
set such that $x_i=1$. 
We have now shown that $\gind\leq\gq\leq\gind -1$. Thus, since 
$\gind$ and $\gq$ are within a constant factor of each other and $\gind$ cannot
be approximated to with a factor of $n^{1-\eps}$ for any constant $0<\eps<1$, 
$\gq$ cannot be approximated to within a factor of $c\,n^{1-\eps}$, or 
$c\,m^{1-\eps}$, or $c'\,\kappa\log^{1-\eps} d_n$ for some positive constants $c$ and $c'$.
\end{proof}

\subsection{Approximation Hardness Results for \MSSBO}
\label{sec-justify-multislot}

A first natural approach to prove an approximation hardness result for \MSSBO\  would be to generalize the approximation hardness result for
the single-slot case {\bf (Q1)} in Theorem~\ref{LB} to the multi-slot case {\bf (Q2)}. 
This can be trivially done by copying the construction of the single-slot case to {\em one} 
of the slots in the multi-slot case. However, after this, one can observe that:

\begin{quote}
{\sf the construction for the single-slot case
cannot again be copied to another slot because of the constraints in Equation~\eqref{multislot-constraint} 
which state that at most one selection variable in each slot can be set to $1$.} 
\end{quote}

\noindent
Formally, the lower bound construction for {\bf (Q1)} can be extended to {\bf (Q2)} as follows:
\begin{itemize}
\item
Identify $y_{i,j,1}$ of {\bf (Q2)} with $y_{i,j}$ of {\bf (Q1)} and set $y_{i,j,2}=y_{i,j,3}=\dots=y_{i,j,s}=0$ in {\bf (Q2)}.

\item
Identify $c_{i,j,1}$ of {\bf (Q2)} with $c_{i,j}$ of {\bf (Q1)} and set $c_{i,j,2}=c_{i,j,3}=\dots=c_{i,j,s}=0$ in {\bf (Q2)}.

\item
Identify $x_{j,k,1}$ of {\bf (Q2)} with $x_j$ of {\bf (Q1)}.
\end{itemize}
This leads to the following approximation hardness result.

\begin{Corollary}
There exist instances of \IMSSBO\ and \FMSSBO, with $n$ keywords, $m=n$ scenarios each with equal probability and $s$ slots, 
such that, unless $\mathsf{ZPP}\!=\!\NP$, {\em any} polynomial-time algorithm for solving these problems must have an approximation ratio 
of $\Omega\left(n^{1-\eps}\right)$ or $\Omega\left(\kappa\,\log^{1-\eps} d_n\right)$, where $0<\eps<1$ is any constant.
\end{Corollary}

The theorem below shows that \IMSSBO\ is {\sf MAX-SNP}-hard even when severely restricted.

\begin{Theorem}[Inapproximability of \IMSSBO\ with two slots]~\\
\label{STRONG-THM}
\IMSSBO\ is {\sf MAX-SNP}-hard for $s=2$
even when $\kappa=1$ and $c_{j,k}=1$ for all $j$ and $k$. 
\end{Theorem}

\begin{proof}
We reduce the {\sf MAX-2SAT-3} problem\footnote{Our reduction approach should also work if we start with {\sf MAX-2SAT}-$k$ for any constant $k$.}
to our problem. {\sf MAX-2SAT-3} is defined 
as follows. We are given a collection of $m$ clauses $C_1,C_2,\dots,C_m$ 
over $n$ Boolean variables $z_1,z_2,\dots,z_n$,  
where every clause is a disjunction of exactly two literals and every variable
occurs exactly $3$ times (and, thus, $m=3n/2$). 
The goal is to find an assignment of truth values to variables to satisfy
a maximum number of clauses. This problem was shown to be {\sf MAX-SNP}-hard in~\cite{BK98}. 

Given an instance of {\sf MAX-2SAT-3} we create an instance of \IMSSBO\ (\IE, {\bf (Q2)})
with $s=2$ as follows. 
Every variable $z_j$ corresponds to a keyword $\K_j$ with two slots.
The variables $x_{j,1}$ and $x_{j,2}$ encode the truth assignments of the variable $z_j$ 
with $x_{j,1}=1$ indicating that $z_j$ is true and $x_{j,2}=1$ indicating 
that $z_j$ is false; we will say that $x_{j,1}$ and $x_{j,2}$ are the slots
corresponding to the literals $z_j$ and $\neg z_j$, respectively. 
There are exactly $m$ scenarios, each with probability $\frac{1}{m}$, defined
in the following manner:
\begin{itemize}
\item
$B_i=1$ for $1\leq i\leq m$. 

\item
$c_{j,k}=1$ for $1\leq j\leq n$ and $1\leq k\leq 2=s$. 

\item
For the $i$th clause $C_i$ containing two {\em literals},
we have the $i$th scenario of the following form. Let $x_{j,k}$ and $x_{j',k'}$ 
be the slots corresponding to the two literals of the clause.
Then we set $y_{i,j,k}=y_{i,j',k'}=1$, and $y_{i,j,k}=0$ if $j\neq j'$ or $k\neq k'$. 
For example, if $C_i=z_2\vee (\neg z_3)$ then 
$y_{i,2,1}=y_{i,3,2}=1$ and $y_{i,j,k}=0$ for all other $j$ and $k$.
\end{itemize}
An inspection of the construction reveals that it satisfies the following:
\begin{itemize}
\item
Because this is an instance of \IMSSBO, by Equation~\eqref{multislot-constraint}, for every $1\leq j\leq n$, 
either $x_{j,1}=1$ or $x_{j,2}=1$ but {\em not both}. 
On the other hand, it is always possible to set at least one of the two variables 
$x_{j,1}=1$ or $x_{j,2}=1$ without decreasing the total payoff.
Thus setting these variables correspond to a truth assignment.

\item
A scenario contributes a payoff of $1$ if and only if {\em at least} one of two slots 
have been selected. Thus, contribution of a scenario correspond to satisfying a clause.
\end{itemize}
By the above observations, we satisfy $m'$ clauses if and only if the above instance
of \IMSSBO\ has a total payoff of $m'$.
\end{proof}

\section{Other Results}
\label{other-results}

\subsection{Improved Algorithms for Special Cases of \SSBO\ and \MSSBO} 
\label{special-cases}

By the phrase ``{\tt within an additive error of $\delta$}'' in 
Lemma~\ref{MFIXED} we
mean that if our solution returns an objective value of $x$ when the optimal value is
$y$ then $|x-y|\leq\delta$.

\begin{Lemma}\label{MFIXED}$\,$
\begin{description}
\item[(a) (Fixed number of scenarios)]
If $m$ is fixed, 
\FMSSBO\ admits 
a {\bf pseudo-polynomial} time 
solution with an absolute error of $\delta$ for any fixed $\delta>0$,
\ISSBO\ admits a 
{\bf pseudo-polynomial} time 
$O(1)$-approximation and 
\IMSSBO\ admits a 
{\bf pseudo-polynomial} time 
$O(\log^2 n)$-approximation.

\item[(b) (Fixed number of keywords)]
If $ns$ is fixed, 
then 
\FMSSBO\ admits 
a {\bf pseudo-polynomial} time 
solution with an absolute error of $\delta$ for any fixed $\delta>0$.

\item[(c) (Logarithmic number of keywords)]
if $ns=O(\log m)$ then 
\IMSSBO\ admits 
a polynomial time exact solution.  

\item[(d) (Fixed number of scenarios and polynomial bids)]
If $m$ is fixed and the maximum size of all the numbers, namely 
$\displaystyle\max\left\{\,\max_{i,j,k} \{y_{i,j,k}\},\,\max_i\{B_i\},\,\max_i \left\{\frac{1}{\eps_i}\right\},\,\max_{i,j,k} \{c_{i,j,k}\}\,\right\}$,
is at most $\poly(n)$
then 
\IMSSBO\ admits 
a polynomial time 
solution with an absolute error of $\delta$ for any fixed $\delta>0$.
\end{description}
\end{Lemma}

\begin{proof}~\\
\noindent
{\bf (a) and (b)}
We prove part~{\bf (a)} as follows (the proof for part~{\bf (b)} is similar).
Consider the \FMSSBO\ problem; 
let $\displaystyle y=\max_{\substack{1\leq i\leq m \\ 1\leq j \leq n \\ 1\leq k\leq s} } \{y_{i,j,k}\}$.

\begin{Proposition}\label{FIXED-SCALE}
Let 
$\balpha^*=(\alpha_1^*,\alpha_2^*,\dots,\alpha_m^*)$ 
and 
$\x^*=(x_{1,1}^*,\dots,x_{1,s}^*,x_{2,1}^*,\dots,x_{2,s}^*,\cdots\cdots,x_{n,1}^*,\dots,x_{n,s}^*)$ 
be the solution vectors for an optimal solution 
of value $\Ex[{\rm payoff}\,^*]=\sum_{i=1}^m\alpha_i^*\left(\sum_{j=1}^n\sum_{k=1}^s y_{i,j,k}\,x_{j,k}^*\right)$. 
Suppose that we approximate the vector $\balpha^*$ by a vector 
$\balpha_\eps=(\alpha_{1,\eps},\dots,\alpha_{m,\eps})$ 
such that $\pmb{|}\,\alpha_i^*-\alpha_{i,\eps}\,\pmb{|}\leq\eps$ for each $i$.
Then, if $\eps\leq\dfrac{\delta}{nsy}$ we can compute a solution
with a total expected payoff of at least
$\Ex[{\rm payoff}]-\delta$.
\end{Proposition}

\begin{proof}
Our algorithm is simple.
Plugging the values of this $\balpha_\eps$ in 
{\bf (Q2)} reduces it to a linear program, which can be solved optimally in polynomial time 
giving a solution vector, say 
$\x_\eps$. Our solution vectors are $\balpha_\eps$ and $\x_\eps$. 
Obviously, all the constraints are satisfied, so we just need to check the total expected payoff of our solution.
For notational convenience, let 
$\F(\balpha,\x)=\sum_{i=1}^m\alpha_i\left(\sum_{j=1}^n\sum_{k=1}^s y_{i,j,k}\,x_{j,k}\right)$
for two vectors 
$\x=(x_{1,1},\dots,x_{1,s},x_{2,1},\dots,x_{2,s},\cdots\cdots,x_{n,1},\dots,x_{n,s})$
and $\balpha=(\alpha_1,\dots,\alpha_m)$; 
thus $\F(\balpha^*,\x^*)=\Ex[{\rm payoff}\,^*]$.
Then,
\begin{gather*}
\pmb{|}\,\F(\balpha^*,\x^*) - \F(\balpha_\eps,\x^*)\,\pmb{|} \leq 
\eps\,\sum_{j=1}^n\sum_{k=1}^s y_{i,j,k}
\leq \eps\,n\,s\,y
\\
\Longrightarrow\,
\F(\balpha_\eps,\x_\eps) 
\geq
\F(\balpha_\eps,\x^*)
\geq
\F(\balpha^*,\x^*) - \eps\,n\,s\,y
\geq
\F(\balpha^*,\x^*) - \delta
\end{gather*}
\qedhere
\end{proof}

To get such a $\balpha_\eps$, 
for every $\alpha_{i,\eps}$ we try out all rational
numbers between $0$ and $1$ of the form
$\frac{j\,\delta}{2\,n\,s\,y}$
for $j=0,1,\dots,\dfrac{2\,n\,s\,y}{\delta}$ until we succeed.
The total number of 
choices is at most 
$\left(\frac{2nsy}{\delta}+1\right)^m$, 
which is {\bf pseudo-polynomial}\footnote{The running time is {\em not} strongly polynomial 
since the input size depends polynomial on $\log_2 y$ (see Section~\ref{sec-notations}).}
in the size of the input since $m$ is fixed. 

The result for \IMSSBO\ follows by using the above proof
with Lemma~\ref{translate}.

\vspace*{0.2in}
\noindent
{\bf (c)}
When $ns=O(\log m)$ then we can try out all
possible $\poly(m)$ assignments of keywords to slots.
For each assignment, we can directly calculate
the values of $\alpha_1,\alpha_2,\ldots,\alpha_m$.
We take the best of all such solutions.

\vspace*{0.2in}
\noindent
{\bf (d)}
Let $p_1(n)$ be a polynomial in $n$ such that 
$\displaystyle\max\left\{\,y,\,\max_i \{B_i\}\mathrel{,}\,\max_{i,j,k}\{w_{i,j,k}\}\,\right\}<p_1(n)$. 
By the proof in part {\bf (a)}, 
to ensure an absolute error of $\delta$, 
it suffices to try all vectors 
$\balpha=(\alpha_1,\alpha_2,\dots,\alpha_m)$ 
in which each $\alpha_i$ is a non-negative rational number with numerator
and denominator at most $p_2(n)$ for some polynomial $p_2(n)$, 
and provide a solution of \IMSSBO\ for this $\balpha$ in polynomial time.
We will refer to $B_i$ as the ``expected budget'' for the $i$th scenario. 
Let $\Ex[\mbox{payoff}\,(j,k,b_1,\dots,b_m)\,]$ 
be the optimal value of the expected payoff when {\em no} slot was selected after the $k$th slot of the $j$th keyword
and the expected budget for the $i$th scenario was $b_i$. 
It is easy to see that the following recurrence holds:
\begin{multline*}
\Ex\left[\,\mbox{payoff}(j,k,b_1,\dots,b_m)\,\right] = \max\Bigg\{\sum_{i=1}^m y_{i,j,k} \,\,+ 
\Ex\left[\,\mbox{payoff}(j-1,s,b_1-\alpha_1 w_{1,j,k},\dots,b_m-\alpha_m w_{m,j,k})\,\right], \\
\Ex\left[\,\mbox{payoff}(j,k-1,b_1,\dots,b_m)\,\right]\,\Bigg\}
\end{multline*}
Based on the above recurrence, it is easy to design a polynomial time dynamic programming
algorithm to compute the optimal solution 
$\Ex\left[\,\mbox{payoff}(n,s,B_1,\dots,B_m)\,\right]$
of \IMSSBO.
\end{proof}

\subsection{Limitations of the Semidefinite Programming Relaxation Approaches for \SSBO}
\label{a1}

\begin{figure}[htbp]
\begin{center}
\begin{tabular}{l}\toprule
{\bf (* Vector program (V) *)} \\
maximize $\sum_{i=1}^m \sum_{j=p}^q y_{i,j}\,\u_{\,i}\centerdot \v_j$ \\
subject to $\forall\, 1\leq i\leq m\colon \,\, \sum_{j=1}^n c_{i,j}\,y_{i,j}\,\u_{\,i}\centerdot \v_j \leq B_i$ \\
\hspace*{0.7in} $\forall\, 1\leq i\leq m\colon\,\,\forall\,1\leq j\leq n\colon\,\,\, \u_{\,i}\centerdot\v_j \geq 0$ \\
\hspace*{0.7in} $\forall\, 1\leq i\leq m\colon\,\,\u_{\,i} \centerdot \u_{\,i} \leq 1$ \\
\hspace*{0.7in} $\forall\, 1\leq i\leq m\colon\,\,\u_{\,i} \in \R^{m+n}$ \\
\hspace*{0.7in} $\forall\,1\leq j\leq n\colon\,\,\, \v_j\centerdot\v_j \leq 1$ \\
\hspace*{0.7in} $\forall\,1\leq j\leq n\colon\,\,\, \v_j\in \R^{m+n}$ \\
\bottomrule
\end{tabular}
\end{center}
\caption{\label{V}{\sf SDP}-relaxation of {\bf (Q1)}.}
\end{figure}

A natural Semidefinite programming ({\sf SDP}) relaxation approach to solve quadratic programs such as {\bf (Q1)}, extensively 
used in existing literatures for efficient approximations of quadratic programs for {\sf MAX}-{\sf CUT},  {\sf MAX}-{\sf 2SAT} and many other problems~\cite{V01}, is as follows.
We first add some redundant inequalities to {\bf (Q1)}.
For every $i$ and $j$ we add the inequality 
$\alpha_i x_j\geq 0$. Clearly, this does not change 
the solutions of {\bf (Q1)}. 
Then, {\bf (Q1)} can be relaxed to a vector program {\bf (V)} by replacing the variables by 
$(m+n)$-dimensional vectors and the product of variables by the inner product (denoted by $\centerdot$) of
the corresponding vectors. The resulting vector program is shown in Fig.~\ref{V}; 
it is well known that {\bf (V)} is a relaxation of {\bf (Q1)} (\EG, see~\cite{V01}).

Since the lower bounds in Theorem~\ref{LB}  
have $\eps<1$ and thus leaves a ``very small'' gap between this lower bound 
and the upper bound in Theorem~\ref{KNAPSACK}, one might wonder 
if the gap can be somewhat narrowed down by designing an approximation algorithm based on the {\sf SDP}-relaxation approaches
whose approximation ratio is, say, $o\left(\frac{m}{\log m}\right)$ or $o\left(\frac{\log d_n}{\log\log d_n}\right)$?
However, we show that the large integrality gap of the {\sf SDP}-relaxation does not allow for such a possibility.

\begin{Lemma}[Limitations of {\sf SDP}-relaxation approaches]\label{SDP}
Let $\kappa=1$.
Let OPT$_{\rm Q1}$ and OPT$_{\rm V}$ be the total optimal payoff for an instance
of {\bf (Q1)} and the optimal value of the objective function of {\bf (V)}, 
respectively. Then,
$\dfrac{\mbox{OPT$_{\rm V}$}}{\mbox{OPT$_{\rm Q1}$}}\geq\dfrac{m}{2}=\Theta\left(\dfrac{\log d_n}{\log\log d_n}\right)$. 
\end{Lemma}

\begin{proof}
We reuse the notations and terminologies used in the proof of Theorem~\ref{LB}.
Let the given graph $G$ be a completely connected graph; 
thus $\gind=1$. 
We construct an instance of \SSBO\ as in Theorem~\ref{LB}. 
Thus, $\gq<1+\gind=2$.
Note that $c_n=d_n=m^{6m}$ and thus $m=\Theta(\log d_n/\log\log d_n)$. 

However, we show that 
OPT$_{\rm vector}\geq m$.
Let $\u_{\,1},\dots,\u_{\,m}$ 
be a set of mutually orthogonal {\em unit-norm} vectors in 
$\R^{m+n}$ and let $\v_i=\u_{\,m-i+1}$ for $1\leq i\leq m$. Thus,
$\u_{\,i}\centerdot\v_j$ is $1$ if $i+j=m+1$ and is $0$ otherwise,
and $\u_{\,i}\centerdot\u_{\,i}=\v_i\centerdot\v_i=1$ for all $i$. 
Obviously, 
$\sum_{i=1}^m \sum_{j=1}^n y_{i,j}\,\u_{\,i}\centerdot\v_j=m$. 
We now verify that this is indeed a valid solution of {\bf (V)} by checking
that it satisfies all the constraints 
$\left(\sum_{j=1}^n w_{i,j}\,\u_{\,i}\centerdot \v_j\right)\leq B_i$ for $1\leq i\leq n$.
It can be seen that 
$\left(\sum_{j=1}^n w_{i,j}\,\u_{\,i}\centerdot\v_j\right)=w_{i,m-i+1}=c_{m-i+1}=B_i$.
\end{proof}

\subsection{Combinatorial Dual of \SSBO\ Problems}

\begin{figure}[htbp]
\begin{center}
\begin{tabular}{l}\toprule
(* Quadratic program {\bf (Dual-Q1)} *) \\
{\sf minimize} $B$ \\
{\sf subject to} $\sum_{i=1}^m \sum_{j=1}^n \alpha_i x_j y_{i,j}\geq P$ \\
\hspace*{0.7in} $\forall\, 1\leq i\leq m\colon\,\ \alpha_i\left(\sum_{j=1}^n w_{i,j}\, x_j\right)\leq \eps_i\,B$ \\
\hspace*{0.7in} $\forall\, 1\leq i\leq m\colon\,\ 0\leq \alpha_i\leq 1$ \\
\hspace*{0.7in} $\forall\,1\leq j\leq n\colon\,0\leq x_i \leq 1$ \\
\bottomrule
\end{tabular}
\end{center}
\caption{\label{ds}Quadratic program for \DSSBO.}
\end{figure}

In \DSSBO, the natural combinatorial dual version of \SSBO, we are given a lower
bound, say $P$, on $\Ex[\mbox{payoff}]$. Our goal is to compute the {\em minimum} possible
value of the budget $B$ of the advertiser such that his/her total expected payoff
is {\em at least} $P$. 
The dual version \DMSSBO\ of \MSSBO\ can be defined in a manner analogous to that of \DSSBO.
\DSSBO\ can be reformulated as 
the quadratic program {\bf (Dual-Q1)} shown in Fig.~\ref{ds}. 

Obviously, \DSSBO\ is $\NP$-hard since \SSBO\ is $\NP$-hard. 
For a given required expected profit $\cP$,
let $\cB_\cP$ be the minimum budget that achieves the expected total profit $\cP$.
We define a {\em bi-criteria approximation} for \DSSBO\ in the
following manner: 

\begin{quote}
a $(\delta,\gamma)$-approximation for \DSSBO,
for $\delta,\gamma\geq 1$, is a solution 
that achieves an expected total profit of at least
$\frac{\cP}{\delta}$ with a budget of $\gamma\,\cB_\cP$.
\end{quote}

\begin{Lemma}\label{DUAL-KNAPSACK}~\\
\noindent
{\bf (a) (Inapproximability of \DSSBO\ via inapproximability of \SSBO)} 
\begin{itemize}
\item
If \FSSBO\ cannot be approximated to within a ratio of $\rho>1$ for
some parameter range,
then \DFSSBO\ also cannot be 
approximated to within a ratio of $\rho$ for the same parameter range.

\item
If \ISSBO\ cannot be approximated to within a ratio of $\rho>1$ for
some parameter range,
then \DISSBO\ also cannot be 
approximated to within a ratio of $\frac{\rho}{200\,\ln m}$ for the same parameter range.
\end{itemize}

\noindent
{\bf (b) (Bi-criterion approximation of \DFSSBO\ via \FSSBO)} 
If \FSSBO\ can be approximated to within a ratio of $\rho>1$
for some parameter range,
then \FSSBO\ 
has a $(\rho,1)$-approximation in the same parameter range.
\end{Lemma}

\begin{proof}
Let $\Ex\left[\,\mbox{payoff}^{\,\cB}\,\right]$
be the optimal total expected payoff for \SSBO\ when 
the budget is $\cB$.
For any constant $\Delta>1$, a solution 
of {\bf (Q1)} with a budget of $\cB$ is obviously also a solution of the same instance of {\bf (Q1)}
with a budget of $\Delta\cB$. 
This implies 
$\Ex\left[\mbox{payoff}^{\,\Delta\cB}\,\right]\geq\Ex\left[\,\mbox{payoff}^{\,\cB}\,\right]$.
Let $p=\sum_{i=1}^m\sum_{j=1}^ny_{i,j}$
and $b=\max_{1\leq i\leq m} \left\{\,\sum_{j=1}^n a_{i,j}\,c_{i,j}\,\right\}$; 
note that both $\log_2 p$ and $\log_2 b$ are polynomial in the size
of the input (see Section~\ref{sec-notations}). 

We prove {\bf (a)} by contradiction.
Suppose that some version of \DSSBO\ has a $\rho$-approximation.
Consider an instance of the {\em same} version of \SSBO\ and suppose the budget is $B$. 
We do a binary search in the range of positive integers $[1,p]$ 
in polynomial time with the approximation algorithm for \DSSBO\ to find a $\cP\in[1,p]$ such
that $\cB_{\cP-1}<\rho\, B$ but  $\cB_\cP\geq\rho\, B$.  
Consider this solution of \DSSBO\ and suppose that $\cB^*$ is the 
actual optimal value of the budget corresponding to the total expected payoff $\cP$. 
Thus, $\cB^*\geq\frac{\cB_\cP}{\rho}\geq B$
and $\Ex\left[\,\mbox{payoff}^{\,\cB_\cP}\,\right]\geq\Ex\left[\,\mbox{payoff}^{\,\cB^*}\,\right]\geq\Ex\left[\,\mbox{payoff}^{\,B}\,\right]$.
Suppose that we now divide every $x_i$ by $\rho$. This provides a valid solution of \FSSBO\ 
with a total expected payoff of at least
$\frac{\Ex\left[\,\mbox{payoff}^{\,\cB_\cP}\,\right]}{\rho}$.
By Lemma~\ref{translate}, from this valid solution of \FSSBO\ 
one can obtain a solution of
\ISSBO\ with a total expected payoff of at least
$\frac{\Ex\left[\,\mbox{payoff}^{\,\cB_\cP}\,\right]}{200\,\rho\,\ln m}$.

To prove {\bf (b)}, 
suppose that some version of \SSBO\ with a budget of $\cB$ has a $\rho$-approximation algorithm.
Consider an instance of the same version of \DSSBO\ with a 
requirement of total expected payoff of $\cP$ and 
let $\cB_\cP$ be the value of an optimal budget for this instance.
Since 
$
\left(1-\dfrac{1}{B+1}\right)\Ex\left[\,\mbox{payoff}^{\,B+1}\,\right]
\leq
\Ex\left[\,\mbox{payoff}^{\,B}\,\right]
\leq
\Ex\left[\,\mbox{payoff}^{\,B+1}\,\right]
$,
we do a binary search in the range of positive integers $[1,b]$ 
in polynomial time 
with the $\rho$-approximation algorithm for \SSBO\ to find a $\cB\in[1,b]$ such
that 
$\frac{\cP}{\rho}\leq\Ex\left[\,\mbox{payoff}^{\,\cB}\,\right]\leq \rho\,\cP +1$.
Thus, this provides a solution of the \DSSBO\ 
with a total expected payoff of at least $\frac{\cP}{\rho}$ and a budget
of at most $\cB_{\,\cP}$, 
giving the desired $(\rho,1)$-approximation in {\bf (b)}.
\end{proof}

\section{Conclusion}

We have presented the first known approximation algorithms as well as hardness results for 
stochastic budget optimization under the scenario model. The scenario model is natural 
in many areas, and it is particularly apt for internet ad systems. 
We obtained our results by making the connection between these problems and a special case
of bipartite quadratic programs; we exploited this intuition crucially in both approximation algorithms
and hardness proofs. These classes of quadratic programs may have independent applications elsewhere.

Our work shows that there are several instances of parameters where stochastic 
budget optimizations are solvable with reasonable
computational resource even with multiple slots. Our hope is that therefore, 
in practice, one can carefully model particular applications such
as sponsored search, so that the parameters are suitable, and  advertisers can optimize their campaigns more effectively than 
is typically done now by applying some of the algorithms in this paper. 

\section*{Acknowledgements}

We thank the reviewers for their detailed comments which improved both the readability and the technical content of the paper.

\end{document}